\documentclass
[preprint,prd,onecolumn,showpacs,showkeys,nobibnotes,nofootinbib,titlepage]{revtex4}%
\usepackage{amsfonts}
\usepackage{amsmath}
\usepackage{amssymb}
\usepackage{graphicx}
\usepackage{float}
\usepackage{rotating}%
\providecommand{\U}[1]{\protect\rule{.1in}{.1in}}

\ifx\pdfoutput\relax\let\pdfoutput=\undefined\fi
\newcount\msipdfoutput
\ifx\pdfoutput\undefined\else
\ifcase\pdfoutput\else
\ifx\paperwidth\undefined\else
\ifdim\paperheight=0pt\relax\else\pdfpageheight\paperheight\fi
\ifdim\paperwidth=0pt\relax\else\pdfpagewidth\paperwidth\fi
\fi\fi\fi
\begin{document}
\preprint{ }
\title[Kerr Metric in Conformal Gravity]{Kerr Metric, Geodesic Motion, and Flyby Anomaly in Fourth-Order Conformal Gravity}
\author{Gabriele U. Varieschi}
\affiliation{Department of Physics, Loyola Marymount University - Los Angeles, CA 90045,
USA\footnote{Email: Gabriele.Varieschi@lmu.edu}}

\begin{abstract}
In this paper we analyze the Kerr geometry in the context of Conformal
Gravity, an alternative theory of gravitation, which is a direct extension of
General Relativity. Following previous studies in the literature, we introduce
an explicit expression of the Kerr metric in Conformal Gravity, which
naturally reduces to the standard General Relativity Kerr geometry in the
absence of Conformal Gravity effects. As in the standard case, we show that
the Hamilton-Jacobi equation governing geodesic motion in a space-time based
on this geometry is indeed separable and that a fourth constant of
motion---similar to Carter's constant---can also be introduced in Conformal
Gravity. Consequently, we derive the fundamental equations of geodesic motion
and show that the problem of solving these equations can be reduced to one of quadratures.

In particular, we study the resulting time-like geodesics in Conformal Gravity
Kerr geometry by numerically integrating the equations of motion for Earth
flyby trajectories of spacecraft. We then compare our results with the
existing data of the Flyby Anomaly in order to ascertain whether Conformal
Gravity corrections are possibly the origin of this gravitational anomaly.
Although Conformal Gravity slightly affects the trajectories of geodesic
motion around a rotating spherical object, we show that these corrections are
minimal and are not expected to be the origin of the Flyby Anomaly, unless
conformal parameters are drastically different from current estimates.

Therefore, our results confirm previous analyses, showing that modifications
due to Conformal Gravity are not likely to be detected at the Solar System
level, but might affect gravity at the galactic or cosmological scale.

\end{abstract}
\startpage{1}
\endpage{ }
\maketitle
\tableofcontents

\section{\label{sect:introduction}Introduction}

In recent years alternative theories of gravity have become progressively more
popular in the scientific literature due to their ability to account for
astrophysical observations without resorting to dark matter (DM) and dark
energy (DE). In fact, despite recent experimental and observational data
supporting the case for DM (for example, see \cite{Clowe:2006eq},
\cite{Aguilar:2013qda}), there is still no conclusive evidence about the
actual origin of this component of the Universe. Similarly, the observation of
an accelerated expansion of the Universe prompted cosmologists to introduce a
DE component within the framework of General Relativity (GR) and standard cosmology.

On the other hand, alternative theories of gravity (such as MOND
\cite{Milgrom:1983ca}-\cite{Milgrom:1983pn}, TeVeS \cite{Bekenstein:2004ne},
NGT \cite{Moffat:1994hv}, and others) have built the case for a possible
paradigm shift by avoiding the exotic DM/DE components and by introducing
possible modifications to standard gravity (for reviews see
\cite{Mannheim:2005bfa}, \cite{Schmidt:2006jt}, \cite{Clifton:2011jh}, and
references therein). This proposed paradigm shift is not so different from
what Einstein himself did in 1915 by extending Newtonian gravity into GR. For
instance, Einstein's explanation of the anomalies in the rate of precession of
the planet Mercury, using GR instead of Newton's law of gravitation, virtually
eliminated a possible \textquotedblleft dark matter\textquotedblright%
\ component in the Solar System (the proposed existence of an unknown
planet---named Vulcan---between Mercury and the Sun, suggested by Le Verrier
as the possible cause of the observed anomalies).

Following this line of thought, we have analyzed in previous publications
(\cite{Varieschi:2008fc}, \cite{Varieschi:2008va}, \cite{Varieschi:2010xs})
the theory of Conformal Gravity\ (CG), a fourth-order extension of Einstein's
second-order General Relativity, as a possible solution to current
cosmological puzzles, such as DM and DE. In particular, in this paper we want
to study the possible implications for geodesic motion of the stationary,
axially symmetric solution (CG\ extension of the Kerr metric).

In Sect. \ref{sect:conformal}, we will start by reviewing the main results of
Conformal Gravity and the equivalent of the Schwarzschild metric in CG. In
Sect. \ref{sect:kerr}, the main part of our paper, we will study the
equivalent of the Kerr metric in CG and perform a separation of variables in
the related Hamilton-Jacobi equation (similar to the procedure introduced by
B. Carter for the GR\ case). In Sect. \ref{sect:flyby}, we will study the
time-like geodesics in fourth-order Kerr space-time and apply our findings to
possible gravitational anomalies detected in our Solar System. In particular,
the so-called Flyby Anomaly (FA) may be related to the fourth-order Kerr
metric analyzed in this paper.

\section{\label{sect:conformal}Conformal gravity and the static, spherically
symmetric metric}

The German mathematician Hermann Weyl pioneered Conformal Gravity in 1918
(\cite{Weyl:1918aa}, \cite{Weyl:1918ib}, \cite{Weyl:1919fi}) by introducing
the so-called \textit{conformal} or \textit{Weyl tensor}, a combination of the
Riemann tensor $R_{\lambda\mu\nu\kappa}$, the Ricci tensor $R_{\mu\nu
}=R^{\lambda}{}_{\mu\lambda\nu}$, and the Ricci scalar $R=R^{\mu}{}_{\mu}$
(see \cite{Varieschi:2008fc} for full details):%
\begin{equation}
C_{\lambda\mu\nu\kappa}=R_{\lambda\mu\nu\kappa}-\frac{1}{2}(g_{\lambda\nu
}R_{\mu\kappa}-g_{\lambda\kappa}R_{\mu\nu}-g_{\mu\nu}R_{\lambda\kappa}%
+g_{\mu\kappa}R_{\lambda\nu})+\frac{1}{6}R\ (g_{\lambda\nu}g_{\mu\kappa
}-g_{\lambda\kappa}g_{\mu\nu}). \label{eqn2.1}%
\end{equation}
The particular form $C^{\lambda}{}_{\mu\nu\kappa}(x)$ of the Weyl tensor is
invariant under the local transformation of the metric:%
\begin{equation}
g_{\mu\nu}(x)\rightarrow\widehat{g}_{\mu\nu}(x)=e^{2\alpha(x)}g_{\mu\nu
}(x)=\Omega^{2}(x)g_{\mu\nu}(x), \label{eqn2.2}%
\end{equation}
where the factor $\Omega(x)=e^{\alpha(x)}$ represents the amount of local
\textquotedblleft stretching\textquotedblright\ of the geometry. Thus, the
name \textquotedblleft conformal\textquotedblright\ indicates a theory
invariant under all possible local stretchings of the space-time.

This generalization of GR was found to be a fourth-order theory, as opposed to
the standard second-order General Relativity: the field equations contain
derivatives up to the fourth order of the metric, with respect to the
space-time coordinates. Following work done by R. Bach \cite{Bach:1921}, C.
Lanczos \cite{Lanczos:1938}, and others, CG was based on the conformal (or
Weyl) action:%
\begin{equation}
I_{W}=-\alpha_{g}\int d^{4}x\ (-g)^{1/2}\ C_{\lambda\mu\nu\kappa}%
\ C^{\lambda\mu\nu\kappa}, \label{eqn2.3}%
\end{equation}
where $g\equiv\det(g_{\mu\nu})$ and $\alpha_{g}$ is the CG coupling constant.
$I_{W}$ is the unique general coordinate scalar action that is completely
locally conformal invariant. Bach \cite{Bach:1921} also introduced the CG
field equations, in the presence of an energy-momentum tensor\footnote{We
follow here the convention \cite{Mannheim:2005bfa}\ of introducing the
energy-momentum tensor $T_{\mu\nu}$ so that the quantity $cT_{00}$ has
dimensions of an energy density.} $T_{\mu\nu}$:%
\begin{equation}
W_{\mu\nu}=\frac{1}{4\alpha_{g}}\ T_{\mu\nu}, \label{eqn2.4}%
\end{equation}
similar to Einstein's equations,%
\begin{equation}
G_{\mu\nu}=R_{\mu\nu}-\frac{1}{2}g_{\mu\nu}\ R=-\frac{8\pi G}{c^{3}}%
\ T_{\mu\nu}. \label{eqn2.5}%
\end{equation}

The \textquotedblleft Bach tensor\textquotedblright\ $W_{\mu\nu}$ in Eq.
(\ref{eqn2.4})\ is analogous to Einstein's curvature tensor $G_{\mu\nu}$, on
the left-hand side of Eq. (\ref{eqn2.5}), but has a much more complex
structure, being defined as:%
\begin{align}
W_{\mu\nu}  &  =-\frac{1}{6}g_{\mu\nu}\ R^{;\lambda}{}_{;\lambda}+\frac{2}%
{3}R_{;\mu;\nu}+R_{\mu\nu}{}^{;\lambda}{}_{;\lambda}-R_{\mu}{}^{\lambda}%
{}_{;\nu;\lambda}-R_{\nu}{}^{\lambda}{}_{;\mu;\lambda}\label{eqn2.6}\\
&  +\frac{2}{3}R\ R_{\mu\nu}-2R_{\mu}{}^{\lambda}\ R_{\lambda\nu}+\frac{1}%
{2}g_{\mu\nu}\ R_{\lambda\rho}\ R^{\lambda\rho}-\frac{1}{6}g_{\mu\nu}%
\ R^{2},\nonumber
\end{align}
and including derivatives up to the fourth order of the metric with respect to
space-time coordinates.

In 1989, Mannheim and Kazanas (\cite{Mannheim:1988dj}, \cite{Kazanas:1988qa})
derived the exact and complete exterior solution for a static, spherically
symmetric source in CG, i.e., the fourth-order analogue of the Schwarzschild
exterior solution in GR. Other exact solutions, such as the equivalent
Reissner-Nordstr\"{o}m, Kerr, and Kerr-Newman in CG, were also derived later
by Mannheim and Kazanas \cite{Mannheim:1990ya} (MK solutions in the following).

The MK solution for a static, spherically symmetric source, in the case
$T_{\mu\nu}=0$ (exterior solution), takes the form
\begin{equation}
ds^{2}=-B(r)\ c^{2}dt^{2}+\frac{dr^{2}}{B(r)}+r^{2}(d\theta^{2}+\sin^{2}%
\theta\ d\phi^{2}), \label{eqn2.7}%
\end{equation}
with%
\begin{equation}
B(r)=1-3\beta\gamma-\frac{\beta(2-3\beta\gamma)}{r}+\gamma r-\kappa r^{2}.
\label{eqn2.8}%
\end{equation}

The parameters in Eq. (\ref{eqn2.8}) are defined as follows: $\beta=\frac
{GM}{c^{2}}\ (%
\operatorname{cm}%
)$ is the geometrized mass, where $M$ is the mass of the (spherically
symmetric) source and $G$ is the universal gravitational constant; two other
parameters, $\gamma\ (%
\operatorname{cm}%
^{-1})$ and $\kappa\ (%
\operatorname{cm}%
^{-2})$, are required by CG. The standard Schwarzschild solution is recovered
as $\gamma,\kappa\rightarrow0$ in the equations above. These two new
parameters are interpreted by MK \cite{Mannheim:1988dj} in the following way:
$\kappa$, and the corresponding term $-\kappa r^{2}$, indicates a background
De Sitter space-time, which is important only over cosmological distances,
since $\kappa$ has a very small value. Similarly, $\gamma$ measures the
departure from the Schwarzschild metric at smaller distances since the $\gamma
r$ term becomes significant over galactic distance scales.

For values of $\gamma\sim10^{-28}-10^{-30}\
\operatorname{cm}%
^{-1}$ the standard Newtonian $\frac{1}{r}$ term dominates at smaller
distances so that CG yields the same results for geodesic motion as those of
standard GR at the scale of the Solar System. At larger galactic distances,
the additional $\gamma r$ term might explain the flat galactic rotation curves
without the need of dark matter (\cite{Mannheim:2005bfa},
\cite{Mannheim:1992vj}, \cite{Mannheim:1996rv}). At even bigger distances, the
quadratic term $-\kappa r^{2}$ plays a role in the dynamics of stars rotating
at the largest possible distances from galactic centers, as recently
established by Mannheim and O'Brien (\cite{Mannheim:2010ti},
\cite{Mannheim:2010xw}, \cite{2012MNRAS.421.1273O}, \cite{Mannheim:2012qw}).

An equivalent way to illustrate the interplay of the different terms in Eq.
(\ref{eqn2.8}) is to write a classical gravitational potential (multiplying by
$c^{2}/2$ all the terms that contain gravitational parameters in that
equation):%
\begin{equation}
V(r)=\frac{c^{2}}{2}\left[  -3\beta\gamma-\frac{\beta(2-3\beta\gamma)}%
{r}+\gamma r-\kappa r^{2}\right]  , \label{eqn2.9}%
\end{equation}
or by considering the potential energy $U(r)=mV(r)$, for a body of mass $m$ in
the gravitational field due to the source mass $M=\frac{\beta c^{2}}{G}$. From
this potential energy we can obtain the equivalent classical central force:%
\begin{equation}
F(r)=-\frac{dU}{dr}=-m\frac{dV}{dr}=-\frac{GMm(2-3\beta\gamma)}{2r^{2}}%
-\frac{mc^{2}\gamma}{2}+mc^{2}\kappa r, \label{eqn2.10}%
\end{equation}
where the right-hand side of the equation reduces to the standard attractive
Newtonian force for $\gamma,\kappa\rightarrow0$. The two additional force
terms, due to CG, represent respectively a constant attractive force and a
linear repulsive force, which might be the origin of the almost flat galactic
rotation curves (due mainly to the constant force term) and of the accelerated
expansion of the Universe (attributable to the repulsive linear term).

In line with this simplified, classical approach to CG, by equating the
magnitude of the force $F(r)$ to the test body centripetal force $mv^{2}/r$,
we readily obtain the circular velocity expression in CG:%
\begin{equation}
v(r)=\sqrt{\frac{GM(2-3\beta\gamma)}{2r}+\frac{c^{2}\gamma r}{2}-c^{2}\kappa
r^{2}}, \label{eqn2.10.1}%
\end{equation}
where, in addition to the Newtonian $1/r$ term, two other conformal terms
appear, linear and quadratic in $r$, which can be used to model galactic
rotation curves in Conformal Gravity.

Extensive data fitting of galactic rotation curves has been carried out by
Mannheim et al. (\cite{Mannheim:2005bfa}, \cite{Mannheim:1992vj},
\cite{Mannheim:1996rv}, \cite{Mannheim:2010ti}, \cite{Mannheim:2010xw},
\cite{2012MNRAS.421.1273O}, \cite{Mannheim:2012qw}) without any DM
contribution. The values of the conformal parameters were determined as
follows \cite{Mannheim:2005bfa}:%
\begin{equation}
\gamma^{\ast}=5.42\times10^{-41}%
\operatorname{cm}%
^{-1},\ \gamma_{0}=3.06\times10^{-30}%
\operatorname{cm}%
^{-1},\ \kappa=9.54\times10^{-54}%
\operatorname{cm}%
^{-2}, \label{eqn2.11}%
\end{equation}
where the CG$\ \gamma$ parameter is split into a $\gamma^{\ast}$, due to the
contributions to the gravitational potential of the individual stars in the
galaxy being studied (stars of reference mass equal to the solar mass
$M_{\odot}$), and a cosmological parameter $\gamma_{0}$,\ due to the
contributions of all the other galaxies in the Universe. The third parameter
$\kappa$\ is considered to be of purely cosmological origin.

By performing a \textquotedblleft kinematical approach\textquotedblright\ to
CG, in previous publications (\cite{Varieschi:2008fc}, \cite{Varieschi:2008va}%
), we have shown a different way to compute the CG\ parameters, obtaining
values which differ by a few orders of magnitude from those in Eq.
(\ref{eqn2.11}):%
\begin{equation}
\gamma=1.94\times10^{-28}%
\operatorname{cm}%
^{-1},\ \kappa=6.42\times10^{-48}%
\operatorname{cm}%
^{-2}. \label{eqn2.12}%
\end{equation}
Using either the values in Eq. (\ref{eqn2.11}) or those in Eq. (\ref{eqn2.12}%
), inside Eqs. (\ref{eqn2.7})-(\ref{eqn2.10.1}), one always obtains very small
corrections to the dynamics of celestial bodies moving within the Solar
System. Therefore, CG corrections were always considered negligible
\cite{Mannheim:2007ug} at the Solar System level, not affecting in any
measurable way the motion of planets, satellites, etc. The effect of CG would
only be appreciable at the galactic and cosmological scales, where
CG\ effectively replaces DM and DE.

One could argue that since the CG potential in Eq. (\ref{eqn2.9}) contains
both a linear and a quadratic term in $r$, the effects of distant stars,
galaxies, etc., should be manifest in our Solar System, thus giving us a way
to confirm or rule out Conformal Gravity as a viable gravitational theory.
However, CG\ effects from distant sources only show themselves as
\textquotedblleft tidal forces\textquotedblright\ on solar-system bodies, and
these tidal effects are completely undetectable in our Solar System
(\cite{Mannheim:2007ug}, \cite{Mannheim:2011p}).

The above analysis suggests that CG\ corrections to the dynamics of the Solar
System are essentially negligible and that CG effects are of importance only
over larger distance scales. However, these considerations were only based on
the static, spherically symmetric CG metric, while many gravitational sources
in the Universe are rotating bodies that might even possess electric/magnetic charge.

In the rest of this paper we will concentrate our efforts on the Kerr metric
in CG, i.e., on the case of a stationary, axially symmetric rotating system
without any charge. The study of solutions for charged bodies goes beyond the
purpose of the current work, also because astrophysical black holes are
thought to be neutral to a very good approximation.

We conclude this section with some general comments about the relation between
CG and GR solutions. It is easy to show that any exterior vacuum solution to
Einstein's gravity is a vacuum solution to CG (\cite{Mannheim:1988dj},
\cite{Mannheim:1990ya}), and thus that any vacuum solution in GR is conformal
to a vacuum solution in CG. On the contrary, there is no general proof that
the converse is also true \cite{Mannheim:2014p}, namely that any vacuum
solution in CG is conformal to a vacuum solution in GR, although this happens
to be the case at least for the static, spherically symmetric CG solution in
Eqs. (\ref{eqn2.7})-(\ref{eqn2.8}), as originally discovered by Mannheim and
Kazanas \cite{Mannheim:1988dj}.

In the literature, there is no similar connection between GR and CG solutions,
in the case of a rotating source \cite{Mannheim:2014p}. This is also due to
the fact that the most general vacuum solution for rotating sources in CG is
not known, since the uniqueness of the fourth-order Kerr solution has not been
established yet. Therefore, the study of the exterior Kerr solution in CG is
fully justified, although it might be argued that `global' vacuum solutions in
CG (as in the case of black holes, where the interior solution is causally
disconnected due to the presence of the event horizon) might be truly
conformally equivalent to their GR counterparts.

The situation is different if interior solutions are also considered, since in
the interior regions GR and CG solutions are not conformal to each other
\cite{Mannheim:1992tr}\ (the former is a solution to a second-order Poisson
equation, while the latter is a solution to a fourth-order Poisson equation).
Therefore, it is the matching of the exterior solution to the interior one, at
the source surface, that fixes the conformal factor of the exterior solution,
thus forcing the CG\ exterior solution to belong to a conformal sector which
is not equivalent, in general, to the one to which the GR exterior solution
belongs. Hence, again, the study of the exterior Kerr solution in CG\ is fully
warranted, especially in the case analyzed later in Sect. \ref{sect:flyby},
where the fourth-order Kerr solution will be used to model the external
gravitational field of the Earth.

\section{\label{sect:kerr}Kerr metric in conformal gravity}

The solutions to the Reissner-Nordstr\"{o}m, Kerr, and Kerr-Newman problems in
CG were first introduced by Mannheim and Kazanas in their 1991 paper
\cite{Mannheim:1990ya}. In particular, considering only the Kerr metric, the
general line element for this geometry can be written as ($c=1$ in the
following):%
\begin{align}
ds^{2}  &  =A(x,y)\ dx^{2}+2E(x,y)\ dx\ dy+C(x,y)\ dy^{2}\label{eqn3.1}\\
&  +D(x,y)\ d\phi^{2}+2F(x,y)\ d\phi\ dt-B(x,y)\ dt^{2},\nonumber
\end{align}
where the coordinates $x$ and $y$ can be identified respectively with the
radial coordinate $r$ and $\cos\theta$.

Following the formalism introduced by B. Carter \cite{1973bhld.book.....D},
the metric is rewritten as \cite{Mannheim:1990ya}:%
\begin{equation}
ds^{2}=(bf-ce)\left[  \frac{dx^{2}}{a}+\frac{dy^{2}}{d}\right]  +\frac
{1}{(bf-ce)}\left[  d\ (b\ d\phi-c\ dt)^{2}-a\ (e\ d\phi-f\ dt)^{2}\right]  ,
\label{eqn3.2}%
\end{equation}
where $a$, $b$, and $c$ are functions depending only on $x\equiv r$, while
$d$, $e$, and $f$ are functions depending only on $y\equiv\cos\theta$. Carter
then obtained the standard GR exterior Kerr solution by setting:%
\begin{align}
a(x)  &  =a^{2}-2MGx+x^{2}\ ;\ b(x)=a^{2}+x^{2}\ ;\ c(x)=a\ ;\label{eqn3.3}\\
d(y)  &  =1-y^{2}\ ;\ e(y)=a(1-y^{2})\ ;\ f(y)=1\nonumber
\end{align}
where $M$ is the source mass, while the parameter $a$ (also denoted by $j$ or
$\alpha$ in the literature\footnote{The function $a(x)$ in Eq. (\ref{eqn3.3})
should not be confused with the angular momentum parameter $a$.}) is the
angular momentum parameter of the rotating source, i.e., $a=J/M$, with $J$
being the source angular momentum. Again, the speed of light is set to $1$ and
should not be confused with the function $c$ defined in Eq. (\ref{eqn3.3}).

It is straightforward to transform the metric in Eqs. (\ref{eqn3.2}%
)-(\ref{eqn3.3}), using the equivalences $x\equiv r$ and $y\equiv\cos\theta$,
into the more familiar expression of the Kerr metric in Boyer-Lindquist
coordinates:%
\begin{align}
ds^{2}  &  =-\left[  1-\frac{2\beta r}{r^{2}+a^{2}\cos^{2}\theta}\right]
\ dt^{2}-\left[  \frac{4\beta ra\sin^{2}\theta}{r^{2}+a^{2}\cos^{2}\theta
}\right]  \ dt\ d\phi+\left[  \frac{r^{2}+a^{2}\cos^{2}\theta}{r^{2}-2\beta
r+a^{2}}\right]  \ dr^{2}\label{eqn3.4}\\
&  +\left[  r^{2}+a^{2}\cos^{2}\theta\right]  \ d\theta^{2}+\left[
r^{2}+a^{2}+\frac{2\beta ra^{2}\sin^{2}\theta}{r^{2}+a^{2}\cos^{2}\theta
}\right]  \sin^{2}\theta\ d\phi^{2}.\nonumber
\end{align}
In the literature, the Kerr metric is also expressed in a more compact way by
using the following functions:%
\begin{equation}
\rho^{2}\equiv r^{2}+a^{2}\cos^{2}\theta\ ;\ \Delta\equiv r^{2}-2\beta
r+a^{2}\ ;\ \Sigma^{2}\equiv\left(  r^{2}+a^{2}\right)  ^{2}-a^{2}\Delta
\sin^{2}\theta; \label{eqn3.5}%
\end{equation}
we will use this simplified notation later, in Sect. \ref{sect:separability}.

In the \textit{weak field} ($\beta/r\ll1$) and \textit{slow rotation}
($a/r\ll1$) limit, the Kerr metric can be simplified as follows:%
\begin{equation}
ds^{2}\approx-\left(  1-\frac{2\beta}{r}\right)  \ dt^{2}-\frac{4\beta
a\sin^{2}\theta}{r}\ dt\ d\phi+\left(  1+\frac{2\beta}{r}\right)
\ dr^{2}+r^{2}\ \left(  d\theta^{2}+\sin^{2}\theta\ d\phi^{2}\right)  .
\label{eqn3.6}%
\end{equation}
These conditions are verified for rotating sources such as the Sun, or planets
in our Solar System.

\subsection{\label{sect:Kerr4}Kerr metric in fourth-order gravity}

The MK solution for the Kerr geometry in CG \cite{Mannheim:1990ya} uses the
metric in Eq. (\ref{eqn3.2}) and the same values for the functions $b$, $c$,
$e$, and $f$ in Eq. (\ref{eqn3.3}), but redefines the other two functions, $a$
and $d$, as fourth-degree polynomials:%
\begin{align}
a(x)  &  =a^{2}+ux+px^{2}+vx^{3}-kx^{4},\label{eqn3.7}\\
d(y)  &  =1+r^{\prime}y-py^{2}+sy^{3}-a^{2}ky^{4},\nonumber
\end{align}
where the parameters $u$, $v$, $r^{\prime}$,\footnote{The parameter
$r^{\prime}$ used here should not be confused with the radial coordinate $r$.}
and $s$ need to verify the CG condition:%
\begin{equation}
uv-r^{\prime}s=0. \label{eqn3.8}%
\end{equation}

Mannheim and Kazanas were able to prove that the combination of Eqs.
(\ref{eqn3.2}), (\ref{eqn3.7}), (\ref{eqn3.8}), and the definitions of the
functions $b$, $c$, $e$, and $f$ given in Eq. (\ref{eqn3.3}) represent the
most general, exact solution in CG for the Kerr problem, although they did not
investigate this solution any further and did not prove its uniqueness.

We can analyze in more detail the fourth-order Kerr solution (i.e., the
Mannheim-Kazanas solution described above) by investigating its connections
with the other metrics. One would expect that the fourth-order Kerr solution
should reduce to the standard second-order Kerr solution, for $\gamma
,\kappa\rightarrow0$, analogous to the second-order reduction of the
fourth-order Schwarzschild metric. Similarly, the fourth-order Kerr solution
(for a rotating source) should reduce to the fourth-order Schwarzschild
solution\footnote{Here we are improperly naming the fourth-order CG solutions
as \textquotedblleft fourth-order Kerr,\textquotedblright\ and
\textquotedblleft fourth-order Schwarzschild\textquotedblright\ solutions. All
these solutions were in fact introduced by Mannheim and Kazanas. By using
these names we simply mean the equivalent CG fourth-order solutions for the
Kerr and Schwarzschild\ geometries.} (for a non-rotating source) for
$a\rightarrow0$, just as it does for the second-order solutions. The
parameters $u$, $p$, $v$, $k$, $r^{\prime}$, and $s$ could then be chosen as
appropriate functions of $\beta$, $\gamma$, and $\kappa$ so that the
connections just described are verified.

However (\cite{Mannheim:1990ya}, \cite{Mannheim:2010p}), the CG fourth-order
Kerr solution (rotating source) is actually \emph{not compatible} with the
(non-rotating) CG fourth-order Schwarzschild metric, for $a\rightarrow0$,
while it is possible to reduce it to the second-order Kerr metric for
$\gamma,\kappa\rightarrow0$. Despite this problem, it was shown by MK
(\cite{Mannheim:1990ya}, Sect. V) that, with the following choice of
parameters and functions:%
\begin{align}
a  &  =0\ ;\ p=1\ ;\ r^{\prime}=s=v=0\label{eqn3.9}\\
a(x)  &  =ux+x^{2}-kx^{4}\nonumber\\
d(y)  &  =1-y^{2},\nonumber
\end{align}
the CG fourth-order Kerr solution reduces to a simple second-order
Schwarzschild-de Sitter metric which, under an appropriate conformal
transformation, can be brought to the form of the CG fourth-order
Schwarzschild metric. Thus, the CG fourth-order Kerr solution is conformally
equivalent to the CG fourth-order Schwarzschild metric, for $a=0$, and this is
sufficient to prove that the two solutions are related as expected
\cite{Mannheim:2010p}.

Following these considerations, we have investigated this connection further
in order to obtain an explicit solution for the CG fourth-order Kerr metric in
terms of the conformal parameters $\gamma$ and $\kappa$. Expanding the
analysis in Sect. V of Ref. \cite{Mannheim:1990ya}, with some additional
algebra, it is easy to show that the metric described by the parameters in Eq.
(\ref{eqn3.9}) also requires:%
\begin{align}
u  &  =-\beta(2-3\beta\gamma)\label{eqn3.10}\\
k  &  =\kappa+\frac{\gamma^{2}(1-\beta\gamma)}{(2-3\beta\gamma)^{2}}\nonumber
\end{align}
in order to be conformally equivalent to the MK solution in Eqs.
(\ref{eqn2.7})-(\ref{eqn2.8}).\footnote{Following the discussion in Sect. V of
Ref. \cite{Mannheim:1990ya}, this requires us to set as integration constant
$c=\gamma/(2-3\beta\gamma)$, which defines the transformation $r=\rho/(1-\rho
c)$ between the radial coordinates, $r$ and $\rho$, of the two equivalent
metrics. This choice also sets the function $p(r)=r/(1+cr)$, which determines
the conformal factor $\Omega(r)=p(r)/r$ connecting the two metrics.}

Therefore, the full CG fourth-order solution for the Kerr geometry can be
written as in Eq. (\ref{eqn3.2}), with the functions:%
\begin{align}
a(x)  &  =a^{2}+ux+x^{2}-kx^{4}\ ;\ b(x)=a^{2}+x^{2}%
\ ;\ c(x)=a\ ;\label{eqn3.11}\\
d(y)  &  =1-y^{2}-a^{2}ky^{4}\ ;\ e(y)=a(1-y^{2})\ ;\ f(y)=1\nonumber
\end{align}
with $u$ and $k$ expressed in terms of the original conformal parameters
$\gamma$ and $\kappa$, as in Eq. (\ref{eqn3.10}).\footnote{Due to the
numerical values of $\gamma$ and $\kappa$ in Eq. (\ref{eqn2.11}), or in Eq.
(\ref{eqn2.12}), we have $k\simeq\kappa$ in Eq. (\ref{eqn3.10}). Therefore,
the parameters $k$ and $\kappa$ are practically equivalent but different in
principle and should not be confused.}

A more practical expression for the CG fourth-order Kerr metric can be
obtained by recasting the previous solution into the more familiar
Boyer-Lindquist coordinates. After some algebra, we obtain:%
\begin{align}
ds^{2}  &  =-\left[  1+\frac{ur}{r^{2}+a^{2}\cos^{2}\theta}-k\left(
r^{2}-a^{2}\cos^{2}\theta\right)  \right]  \ dt^{2}\label{eqn3.12}\\
&  +2\left[  \frac{ura\sin^{2}\theta+ka\left(  a^{2}(r^{2}+a^{2})\cos
^{4}\theta-r^{4}\sin^{2}\theta\right)  }{r^{2}+a^{2}\cos^{2}\theta}\right]
\ dt\ d\phi\nonumber\\
&  +\left[  \frac{r^{2}+a^{2}\cos^{2}\theta}{r^{2}+ur+a^{2}-kr^{4}}\right]
\ dr^{2}+\left[  \frac{r^{2}+a^{2}\cos^{2}\theta}{1-ka^{2}\cos^{2}\theta
\cot^{2}\theta}\right]  \ d\theta^{2}\nonumber\\
&  +\left[  \left(  r^{2}+a^{2}-\frac{ura^{2}\sin^{2}\theta}{r^{2}+a^{2}%
\cos^{2}\theta}\right)  \sin^{2}\theta+ka^{2}\frac{\left(  r^{4}\sin^{4}%
\theta-(r^{2}+a^{2})^{2}\cos^{4}\theta\right)  }{r^{2}+a^{2}\cos^{2}\theta
}\right]  \ d\phi^{2},\nonumber
\end{align}
where $u$ and $k$ are again defined through Eq. (\ref{eqn3.10}). Since
$u\rightarrow-2\beta$ and $k\rightarrow0$, for $\gamma,\kappa\rightarrow0$, it
is easy to confirm that the fourth-order Kerr metric in Eq. (\ref{eqn3.12})
correctly reduces to the second-order Kerr metric of Eq. (\ref{eqn3.4}), when
the CG parameters $\gamma$ and $\kappa$ are set to zero. On the other hand,
setting $a=0$, i.e., for a non-rotating source, we obtain from Eq.
(\ref{eqn3.12}) the already mentioned Schwarzschild-de Sitter metric, which is
conformally equivalent to the MK solution in Eqs. (\ref{eqn2.7})-(\ref{eqn2.8}).

Therefore, the metric in Eq. (\ref{eqn3.12}) represents the exact CG
fourth-order solution for the Kerr geometry, expressed explicitly in terms of
the conformal parameters $\gamma$ and $\kappa$. This solution will be used in
the rest of the paper. As in the original study by Mannheim and Kazanas
\cite{Mannheim:1990ya}, we will not attempt to prove that this solution is
also unique.

Just as it was done for the second-order solution, we can also consider the
\textit{weak field} ($u/r\ll1$), \textit{slow rotation} ($a/r\ll1$) limit of
the previous metric (assuming also $kr^{2}\ll1$, i.e., for values of the
radial distance not too large, so that the quantity $kr^{2}$ is small and
comparable with the other two quantities above):%
\begin{align}
ds^{2}  &  \approx-\left(  1+\frac{u}{r}-kr^{2}\right)  \ dt^{2}+2\left(
\frac{u}{r}-kr^{2}\right)  a\sin^{2}\theta\ dt\ d\phi\label{eqn3.13}\\
&  +\left(  1-\frac{u}{r}+kr^{2}\right)  \ dr^{2}+r^{2}\left(  d\theta
^{2}+\sin^{2}\theta\ d\phi^{2}\right)  ,\nonumber
\end{align}
which reduces to the \textit{weak field}, \textit{slow rotation} Kerr
second-order expression in Eq. (\ref{eqn3.6}), for $u\rightarrow-2\beta$ and
$k\rightarrow0$, in the non-conformal gravity case. We can also remark that,
from the asymptotic behaviors (i.e., in the limit $r\rightarrow\infty$) of the
metric coefficients in both Eqs. (\ref{eqn3.12}) and (\ref{eqn3.13}), the
fourth-order Kerr metric does not asymptotically reduce to flat space-time, as
in the second-order Kerr solution. This is, of course, a general feature of
all CG solutions, including the original MK metric for the static, spherically
symmetric source in Eqs. (\ref{eqn2.7}) and (\ref{eqn2.8}).

In the following sections, we will continue to use the full, exact solution in
Eq. (\ref{eqn3.12}), without any approximation. The correctness of this
expression has been tested with the aid of a Mathematica program, which was
developed for other studies in CG \cite{Varieschi:2012ic}. This Mathematica
routine is able to compute and manipulate symbolically all relevant tensors in
both GR and CG. In particular, for a given metric such as the one in Eq.
(\ref{eqn3.12}), our program can compute the full expression of the Bach
tensor $W_{\mu\nu}$ in Eq. (\ref{eqn2.6}) and verify that $W_{\mu\nu}=0$, as
follows from Eq. (\ref{eqn2.4}), for all exterior solutions where $T_{\mu\nu
}=0$. Therefore, our fundamental solution in Eq. (\ref{eqn3.12}) agrees with
the field equations of Conformal Gravity, for a Kerr geometry.

\subsection{\label{sect:separability}The separability of the Hamilton-Jacobi
equation}

Since its introduction in 1963, the Kerr metric \cite{Kerr:1963ud} in Eq.
(\ref{eqn3.4}) has played an essential role in the description of rotating
black holes in General Relativity. Another fundamental theoretical advance in
this field of GR was achieved in 1968 by B. Carter (\cite{1968PhRv..174.1559C}%
, \cite{Carter:1968ks}) who demonstrated the separability of the related
Hamilton-Jacobi equation and deduced the existence of an additional conserved
quantity: Carter's constant $\mathcal{Q}$ (or $\mathcal{K}$).

This discovery essentially solved the problem of geodesic motion in Kerr
space-time: in addition to the energy, the angular momentum about the symmetry
axis, and the norm of the four-velocity, Carter's constant $\mathcal{Q}$
provided a fourth conserved quantity which reduced the problem of solving the
equations of geodesic motion to one involving only quadratures. The complete
analysis of the Kerr space-time, including the separability of the
Hamilton-Jacobi equation and the study of geodesic motion, is presented in
detail in the classic book by S. Chandrasekhar (\cite{1992mtbh.book.....C},
Chapters 6-7). In this section we will draw a parallel between the discussion
by Chandrasekhar (especially the analysis in sections 62 and 64 of Chapter 7)
and the similar discussion of the problem in CG fourth-order Kerr space-time.

We begin by adopting the same notation used in Chandrasekhar's book ($c=G=1$
in the following), which is slightly different from the one used so far, for a
better comparison of the GR and CG\ Kerr space-time formalism. The Kerr metric
in Eqs. (\ref{eqn3.4})-(\ref{eqn3.5}) can be rewritten as follows
(\cite{1992mtbh.book.....C}, Chapter 6):%
\begin{equation}
ds^{2}=-\rho^{2}\frac{\Delta}{\Sigma^{2}}\ dt^{2}+\frac{\Sigma^{2}}{\rho^{2}%
}\left[  d\phi-\frac{2aMr}{\Sigma^{2}}dt\right]  ^{2}\sin^{2}\theta
\ +\frac{\rho^{2}}{\Delta}\ dr^{2}+\rho^{2}d\theta^{2} \label{eqn3.14}%
\end{equation}
with%
\begin{equation}
\rho^{2}\equiv r^{2}+a^{2}\cos^{2}\theta\ ;\ \Delta\equiv r^{2}-2Mr+a^{2}%
\ ;\ \Sigma^{2}\equiv\left(  r^{2}+a^{2}\right)  ^{2}-a^{2}\Delta\sin
^{2}\theta. \label{eqn3.15}%
\end{equation}

In this new notation $a$ still represents the angular momentum parameter
($a=J/M$), while $M$, in Eqs. (\ref{eqn3.14})-(\ref{eqn3.15}), directly
indicates the geometrized mass, i.e., $M\equiv\beta$ ($\beta$ was equal to
$\frac{GM}{c^{2}}$ in previous equations, where $M$ was the non-geometrized
mass). It is easy to check that Eq. (\ref{eqn3.14}) is the same metric of Eq.
(\ref{eqn3.4}). We use a space-like convention for all the metrics in this
paper, while Chandrasekhar adopted a time-like convention for the metric
signature in his book.

Our CG fourth-order Kerr metric in Eq. (\ref{eqn3.12}) can also be recast in
this notation:\footnote{In the form of Eq. (\ref{eqn3.16}) our metric looks
similar to the well-known Kerr-AdS$_{4}$ black hole metric
\cite{1999PhRvD..59f4005H} for an asymptotically anti-de Sitter space.
However, our solution is different and fully satisfies Eq. (\ref{eqn2.4}) of
Conformal Gravity with $T_{\mu\nu}=0$. It should also be noted that, since our
fourth-order Kerr metric is not asymptotically flat, it might be also
superradiantly unstable (as in the case of standard general relativity, see
\cite{Cardoso:2004hs} or \cite{Berti:2009kk} for a review).}%
\begin{equation}
ds^{2}=-\rho^{2}\frac{\widetilde{\Delta}_{r}\ \widetilde{\Delta}_{\theta}%
}{\widetilde{\Sigma}^{2}}\ dt^{2}+\frac{\widetilde{\Sigma}^{2}}{\rho^{2}%
}\left[  d\phi+\frac{\widetilde{\Delta}_{r}-\left(  r^{2}+a^{2}\right)
\widetilde{\Delta}_{\theta}}{\widetilde{\Sigma}^{2}}a\ dt\right]  ^{2}\sin
^{2}\theta\ +\frac{\rho^{2}}{\widetilde{\Delta}_{r}}\ dr^{2}+\frac{\rho^{2}%
}{\widetilde{\Delta}_{\theta}}d\theta^{2}, \label{eqn3.16}%
\end{equation}
with extended definitions for the CG auxiliary quantities and functions:%
\begin{align}
\widetilde{M}  &  \equiv M\left(  1-\frac{3}{2}\ M\gamma\right)
\ ;\ \widetilde{\Delta}_{r}\equiv r^{2}-2\widetilde{M}r+a^{2}-kr^{4}%
\ ;\label{eqn3.17}\\
\widetilde{\Delta}_{\theta}  &  \equiv1-ka^{2}\cos^{2}\theta\cot^{2}%
\theta\ ;\ \widetilde{\Sigma}^{2}\equiv\widetilde{\Delta}_{\theta}\left(
r^{2}+a^{2}\right)  ^{2}-a^{2}\widetilde{\Delta}_{r}\sin^{2}\theta
\ .\ \nonumber
\end{align}
It is easy to check that the CG auxiliary quantities in the last
equation\footnote{In the metric of Eq. (\ref{eqn3.12}) we used the quantity
$u=-\beta(2-3\beta\gamma)$, also defined in Eq. (\ref{eqn3.10}) with
$\beta=GM/c^{2}$. Replacing $\beta$ with the geometrized mass $M$ yields:
$u=-M(2-3M\gamma)=-2M(1-\frac{3}{2}M\gamma)=-2\widetilde{M}$, which explains
the definition of $\widetilde{M}$ in Eq. (\ref{eqn3.17}).} reduce to those in
Eq. (\ref{eqn3.15}) in the non-conformal gravity case, i.e., for
$\gamma,\kappa\rightarrow0$ ($\widetilde{\Delta}_{\theta}\rightarrow1$, in the
non-CG case). Therefore, the CG metric in Eq. (\ref{eqn3.16}) reduces to the
GR metric in Eq. (\ref{eqn3.14}), when the non-conformal case is considered.

It is beyond the scope of this work to analyze in detail the singularities and
other properties of the metric in Eq. (\ref{eqn3.12}) or Eq. (\ref{eqn3.16}).
We just remark that our fourth-order metric is singular for $\widetilde{\Delta
}_{r}=0$, $\widetilde{\Delta}_{\theta}=0$, and $\rho^{2}=0$ in a similar way
to the second-order metric, which is singular for $\Delta=0$ and $\rho^{2}=0$.
A limited analysis of the curvature invariants of the fourth-order Kerr metric
shows that the latter singularity, for $\rho^{2}=0$, is also a curvature
singularity in the fourth-order case, while $\widetilde{\Delta}_{r}\equiv
r^{2}-2\widetilde{M}r+a^{2}-kr^{4}=0$ is likely to be the coordinate
singularity related to the black hole horizon, but with a more complex
structure due to the presence of the $-kr^{4}$ term. An additional
singularity, for $\widetilde{\Delta}_{\theta}\equiv1-ka^{2}\cos^{2}\theta
\cot^{2}\theta=0$, appears in the fourth-order case: since the $k$ parameter
is very small, this singularity corresponds to $\theta\simeq0,\ \pi$. This is
also likely to correspond to a coordinate singularity.

Following the treatment outlined in \cite{1992mtbh.book.....C} (Chapter 7,
\S 62), the Hamilton-Jacobi equation for geodesic motion in a space-time with
metric tensor $g^{\mu\nu}$ is%
\begin{equation}
2\frac{\partial S}{\partial\tau}=g^{\mu\nu}\frac{\partial S}{\partial x^{\mu}%
}\frac{\partial S}{\partial x^{\nu}}, \label{eqn3.18}%
\end{equation}
where $\tau$ is an affine parameter along the geodesic, which can be
identified with proper time for time-like geodesics, and $S$ is Hamilton's
principal function. The Lagrangian\footnote{The negative sign appearing in Eq.
(\ref{eqn3.19}), after the first equality sign, is due to our choice of the
metric signature (different from the one adopted by Chandrasekhar), so that
all the following equations can be compared directly with those in Ref.
\cite{1992mtbh.book.....C}. It should also be noted that the Lagrangian used
to study geodesics is not conformally invariant, even in the non-rotating
black hole case \cite{Said:2014lua}.} $%
\mathcal{L}%
$ related to the metric in Eq. (\ref{eqn3.16}) is:
\begin{align}
2%
\mathcal{L}%
&  =-g_{\mu\nu}\frac{\partial x^{\mu}}{\partial\tau}\frac{\partial x^{\nu}%
}{\partial\tau}=\left[  1-\frac{2\widetilde{M}r}{\rho^{2}}-k\left(
r^{2}-a^{2}\cos^{2}\theta\right)  \right]  \overset{\centerdot}{t}%
^{2}\label{eqn3.19}\\
&  +\frac{4a\widetilde{M}r\sin^{2}\theta-2ka\left[  a^{2}(r^{2}+a^{2})\cos
^{4}\theta-r^{4}\sin^{2}\theta\right]  }{\rho^{2}}\overset{\centerdot
}{t}\overset{\centerdot}{\ \phi}-\frac{\rho^{2}}{_{\widetilde{\Delta}_{r}}%
}\overset{\centerdot}{r}^{2}-\frac{\rho^{2}}{\widetilde{\Delta}_{\theta}%
}\overset{\centerdot}{\theta}^{2}\nonumber\\
&  -\left[  \left(  r^{2}+a^{2}+\frac{2a^{2}\widetilde{M}r\sin^{2}\theta}%
{\rho^{2}}\right)  \sin^{2}\theta+ka^{2}\frac{\left(  r^{4}\sin^{4}%
\theta-(r^{2}+a^{2})^{2}\cos^{4}\theta\right)  }{\rho^{2}}\right]
\overset{\centerdot}{\phi}^{2},\nonumber
\end{align}
where the dot over the variables indicates the derivative with respect to the
affine parameter $\tau$.

The energy and the angular momentum integrals are easily computed:%
\begin{align}
p_{t}  &  =\frac{\partial%
\mathcal{L}%
}{\partial\overset{\centerdot}{t}}=\left[  1-\frac{2\widetilde{M}r}{\rho^{2}%
}-k\left(  r^{2}-a^{2}\cos^{2}\theta\right)  \right]  \overset{\centerdot
}{t}\label{eqn3.20}\\
&  +\frac{2a\widetilde{M}r\sin^{2}\theta-ka\left[  a^{2}(r^{2}+a^{2})\cos
^{4}\theta-r^{4}\sin^{2}\theta\right]  }{\rho^{2}}\overset{\centerdot}{\phi
}\,\nonumber\\
&  =E=constant\nonumber
\end{align}
and%
\begin{align}
-p_{\phi}  &  =-\frac{\partial%
\mathcal{L}%
}{\partial\overset{\centerdot}{\phi}}=\frac{-2a\widetilde{M}r\sin^{2}%
\theta+ka\left[  a^{2}(r^{2}+a^{2})\cos^{4}\theta-r^{4}\sin^{2}\theta\right]
}{\rho^{2}}\overset{\centerdot}{t}\label{eqn3.21}\\
&  +\left[  \left(  r^{2}+a^{2}+\frac{2a^{2}\widetilde{M}r\sin^{2}\theta}%
{\rho^{2}}\right)  \sin^{2}\theta+ka^{2}\frac{\left(  r^{4}\sin^{4}%
\theta-(r^{2}+a^{2})^{2}\cos^{4}\theta\right)  }{\rho^{2}}\right]
\overset{\centerdot}{\phi}\nonumber\\
&  =L_{z}=constant.\nonumber
\end{align}

These last three equations are the CG equivalent of Eqs. (147)-(149) in
Chapter 7 of Ref. \cite{1992mtbh.book.....C}, where $E$ and $L_{z}$ are
interpreted respectively as energy per unit mass and angular momentum---in the
axial direction---per unit mass. The other two canonical momenta%
\begin{align}
p_{r}  &  =\frac{\partial%
\mathcal{L}%
}{\partial\overset{\centerdot}{r}}=-\frac{\rho^{2}}{\widetilde{\Delta}_{r}%
}\overset{\centerdot}{r}\label{eqn3.22}\\
p_{\theta}  &  =\frac{\partial%
\mathcal{L}%
}{\partial\overset{\centerdot}{\theta}}=-\frac{\rho^{2}}{\widetilde{\Delta
}_{\theta}}\overset{\centerdot}{\theta}\nonumber
\end{align}
are obviously non-conserved quantities.

The third integral of motion is related to the conservation of the rest mass,
which can be expressed as the constancy of the norm of the four-velocity
$\mathbf{k}$:%
\begin{equation}
\left\vert \mathbf{k}\right\vert ^{2}=\delta_{1}=%
\genfrac{\{}{\}}{0pt}{}{1\ ;\text{ for time-like geodesics}}{0\ ;\text{ for
null geodesics \ \ \ \ \ \ }}%
. \label{eqn3.23}%
\end{equation}
The fourth integral of motion is obtained by separation of variables in the
Hamilton-Jacobi equation, by seeking a solution to Eq. (\ref{eqn3.18}) of the
form:%
\begin{equation}
S=\frac{1}{2}\delta_{1}\tau-Et+L_{z}\phi+S_{r}(r)+S_{\theta}(\theta),
\label{eqn3.24}%
\end{equation}
where $S_{r}$ and $S_{\theta}$ are functions only of $r$ and $\theta$, respectively.

The procedure for the separation of variables follows closely the one
described in pages 344-347 of Ref. \cite{1992mtbh.book.....C}, although the
related algebra is more challenging, due to the additional terms of CG. For
brevity, we will present here the final results of these algebraic
computations, which were first derived by hand and then checked with our
symbolic Mathematica routines for accuracy.

After various algebraic transformations, the Hamilton-Jacobi equation can be
rewritten as%
\begin{align}
&  \left\{  \widetilde{\Delta}_{r}\left(  \frac{dS_{r}}{dr}\right)  ^{2}%
-\frac{1}{\widetilde{\Delta}_{r}}\left[  \left(  r^{2}+a^{2}\right)
E-aL_{z}\right]  ^{2}+\delta_{1}r^{2}\right\} \label{eqn3.25}\\
+  &  \left\{  \widetilde{\Delta}_{\theta}\left(  \frac{dS_{\theta}}{d\theta
}\right)  ^{2}+\frac{1}{\widetilde{\Delta}_{\theta}}(aE\sin^{2}\theta
-L_{z})^{2}\csc^{2}\theta+\delta_{1}a^{2}\cos^{2}\theta\right\}  =0\nonumber
\end{align}
and the equation can now be separated by introducing a separation constant
$\widetilde{\mathcal{K}}$, which is related to Carter's constant
$\widetilde{\mathcal{Q}}$ as in the standard second-order
analysis:\footnote{As for the other quantities, in the following we will
denote the fourth-order constants $\widetilde{\mathcal{K}}$ (or$\mathcal{\ }%
\widetilde{\mathcal{Q}}$) with a tilde ($\symbol{126}$) superscript to
distinguish them from their second-order equivalents, $\mathcal{K}$
(or$\mathcal{\ Q}$).}%
\begin{equation}
\widetilde{\mathcal{K}}\mathcal{=}\widetilde{\mathcal{Q}}\mathcal{+}%
(L_{z}-aE)^{2}. \label{eqn3.26}%
\end{equation}

By equating the two parts of Eq. (\ref{eqn3.25}) respectively to
$-\widetilde{\mathcal{K}}$ and $+\widetilde{\mathcal{K}}$, we obtain the two
separate equations:%
\begin{align}
\widetilde{\Delta}_{r}^{2}\left(  \frac{dS_{r}}{dr}\right)  ^{2}  &  =\left[
\left(  r^{2}+a^{2}\right)  E-aL_{z}\right]  ^{2}-\widetilde{\Delta}%
_{r}\left(  \widetilde{\mathcal{K}}+\delta_{1}r^{2}\right) \label{eqn3.27}\\
\widetilde{\Delta}_{\theta}^{2}\left(  \frac{dS_{\theta}}{d\theta}\right)
^{2}  &  =-(aE\sin^{2}\theta-L_{z})^{2}\csc^{2}\theta+\widetilde{\Delta
}_{\theta}\left(  \widetilde{\mathcal{K}}-\delta_{1}a^{2}\cos^{2}%
\theta\right)  ,\nonumber
\end{align}
which are similar to Eqs. (172)-(173) in Chapter 7 of Ref.
\cite{1992mtbh.book.....C}, apart from a different arrangement of the terms in
the two equations due to the presence of the additional conformal factor
$\widetilde{\Delta}_{\theta}$.

Setting $\gamma,\kappa=0$, and thus $\widetilde{\Delta}_{r}\rightarrow\Delta$,
$\widetilde{\Delta}_{\theta}\rightarrow1$, $\widetilde{M}\rightarrow M$,
$\widetilde{\mathcal{K}}\rightarrow\mathcal{K}$, etc., we can indeed obtain
the second-order equations of Ref. \cite{1992mtbh.book.....C} by using the
identity $(aE\sin^{2}\theta-L_{z})^{2}\csc^{2}\theta=(L_{z}^{2}\csc^{2}%
\theta-a^{2}E^{2})\cos^{2}\theta+(L_{z}-aE)^{2}$ and reinserting
$\mathcal{Q=K-}(L_{z}-aE)^{2}$ in the equations. Since the fourth-order
equations are more easily written using $\widetilde{\mathcal{K}}$, instead of
$\widetilde{\mathcal{Q}}$, we will adopt this notation in the following.

The solution procedure then continues in a way similar to the second-order
case (see again Ref. \cite{1992mtbh.book.....C}). With the abbreviations:%
\begin{align}
\widetilde{R}(r)  &  \equiv\left[  \left(  r^{2}+a^{2}\right)  E-aL_{z}%
\right]  ^{2}-\widetilde{\Delta}_{r}\left(  \widetilde{\mathcal{K}}+\delta
_{1}r^{2}\right) \label{eqn3.28}\\
\widetilde{\Theta}(\theta)  &  \equiv-(aE\sin^{2}\theta-L_{z})^{2}\csc
^{2}\theta+\widetilde{\Delta}_{\theta}\left(  \widetilde{\mathcal{K}}%
-\delta_{1}a^{2}\cos^{2}\theta\right) \nonumber\\
&  =-\left[  (L_{z}^{2}\csc^{2}\theta-a^{2}E^{2})\cos^{2}\theta+(L_{z}%
-aE)^{2}\right]  +\widetilde{\Delta}_{\theta}\left(  \widetilde{\mathcal{K}%
}-\delta_{1}a^{2}\cos^{2}\theta\right)  ,\nonumber
\end{align}

similar to the second-order expressions (Eqs. (174)-(175) in Chapter 7 of Ref.
\cite{1992mtbh.book.....C}):%
\begin{align}
R(r)  &  \equiv\left[  \left(  r^{2}+a^{2}\right)  E-aL_{z}\right]
^{2}-\Delta\left[  \mathcal{Q+}(L_{z}-aE)^{2}+\delta_{1}r^{2}\right]
\label{eqn3.29}\\
&  =\left[  \left(  r^{2}+a^{2}\right)  E-aL_{z}\right]  ^{2}-\Delta\left(
\mathcal{K}+\delta_{1}r^{2}\right) \nonumber\\
\Theta(\theta)  &  \equiv\mathcal{Q-[}a^{2}(\delta_{1}-E^{2})+L_{z}^{2}%
\csc^{2}\theta\mathcal{]}\cos^{2}\theta\nonumber\\
&  =-\left[  (L_{z}^{2}\csc^{2}\theta-a^{2}E^{2})\cos^{2}\theta+(L_{z}%
-aE)^{2}\right]  +\left(  \mathcal{K}-\delta_{1}a^{2}\cos^{2}\theta\right)
,\nonumber
\end{align}
we can write the solution for $S$, from Eq. (\ref{eqn3.24}), as%
\begin{equation}
S=\frac{1}{2}\delta_{1}\tau-Et+L_{z}\phi+%
{\displaystyle\int\nolimits^{r}}
\frac{\sqrt{\widetilde{R}(r)}}{\widetilde{\Delta}_{r}}dr+%
{\displaystyle\int\nolimits^{\theta}}
\frac{\sqrt{\widetilde{\Theta}(\theta)}}{\widetilde{\Delta}_{\theta}}d\theta.
\label{eqn3.30}%
\end{equation}

To obtain the equations of motion, we then set to zero the partial derivatives
of $S$ with respect to the constants of motion, $\widetilde{\mathcal{K}}$ (or
$\widetilde{\mathcal{Q}}$), $\delta_{1}$, $E$, and $L_{z}$, following again a
procedure similar to the second-order case. We will just report here the final
results for our CG fourth-order case and compare them directly with the
GR\ second-order case (see Eqs. (183)-(186) in Chapter 7 of Ref.
\cite{1992mtbh.book.....C}):%
\begin{equation}
\overset{\centerdot}{r}^{2}=\left(  \frac{dr}{d\tau}\right)  ^{2}=%
\begin{Bmatrix}
\frac{\widetilde{R}(r)}{\rho^{4}}=\frac{\widetilde{\Delta}_{r}{}^{2}}{\rho
^{4}}p_{r}^{2}\ ;\text{ 4th-order}\\
\frac{R(r)}{\rho^{4}}=\frac{\Delta^{2}}{\rho^{4}}p_{r}^{2}\ ;\text{ 2nd-order}%
\end{Bmatrix}
, \label{eqn3.31}%
\end{equation}%
\begin{equation}
\overset{\centerdot}{\theta}^{2}=\left(  \frac{d\theta}{d\tau}\right)  ^{2}=%
\begin{Bmatrix}
\frac{\widetilde{\Theta}(\theta)}{\rho^{4}}=\frac{\widetilde{\Delta}_{\theta
}^{2}}{\rho^{4}}p_{\theta}^{2}\ ;\text{ 4th-order}\\
\frac{\Theta(\theta)}{\rho^{4}}=\frac{1}{\rho^{4}}p_{\theta}^{2}\ ;\text{
2nd-order}%
\end{Bmatrix}
, \label{eqn3.32}%
\end{equation}%
\begin{equation}
\overset{\centerdot}{\phi\ }=\left(  \frac{d\phi}{d\tau}\right)  =%
\begin{Bmatrix}
\frac{1}{\rho^{2}\widetilde{\Delta}_{r}\widetilde{\Delta}_{\theta}}\left\{
\left[  \left(  r^{2}+a^{2}\right)  \widetilde{\Delta}_{\theta}%
-\widetilde{\Delta}_{r}\right]  aE+\left[  \widetilde{\Delta}_{\theta}\rho
^{2}+\widetilde{\Delta}_{r}-\left(  r^{2}+a^{2}\right)  \widetilde{\Delta
}_{\theta}\right]  L_{z}\csc^{2}\theta\right\} \\
=\frac{1}{\rho^{2}}\left\{  \frac{a\left[  \left(  r^{2}+a^{2}\right)
E-aL_{z}\right]  }{\widetilde{\Delta}_{r}}+\frac{(L_{z}\csc^{2}\theta
-aE)}{\widetilde{\Delta}_{\theta}}\right\}  \ ;\text{ 4th-order}\\
\frac{1}{\rho^{2}\Delta}[2aMrE+(\rho^{2}-2Mr)L_{z}\csc^{2}\theta]\\
=\frac{1}{\rho^{2}}\left\{  \frac{a\left[  \left(  r^{2}+a^{2}\right)
E-aL_{z}\right]  }{\Delta}+(L_{z}\csc^{2}\theta-aE)\right\}  \ ;\text{
2nd-order}%
\end{Bmatrix}
, \label{eqn3.33}%
\end{equation}%
\begin{equation}
\overset{\centerdot}{t\ }=\left(  \frac{dt}{d\tau}\right)  =%
\begin{Bmatrix}
\frac{1}{\rho^{2}\widetilde{\Delta}_{r}\widetilde{\Delta}_{\theta}}\left(
\widetilde{\Sigma}^{2}E+\left[  \widetilde{\Delta}_{r}-\left(  r^{2}%
+a^{2}\right)  \widetilde{\Delta}_{\theta}\right]  aL_{z}\right) \\
=\frac{1}{\rho^{2}}\left\{  \frac{\left(  r^{2}+a^{2}\right)  \left[  \left(
r^{2}+a^{2}\right)  E-aL_{z}\right]  }{\widetilde{\Delta}_{r}}+\frac{a\sin
^{2}\theta(L_{z}\csc^{2}\theta-aE)}{\widetilde{\Delta}_{\theta}}\right\}
\ ;\text{ 4th-order}\\
\frac{1}{\rho^{2}\Delta}\left(  \Sigma^{2}E-2aMrL_{z}\right) \\
=\frac{1}{\rho^{2}}\left\{  \frac{\left(  r^{2}+a^{2}\right)  \left[  \left(
r^{2}+a^{2}\right)  E-aL_{z}\right]  }{\Delta}+a\sin^{2}\theta(L_{z}\csc
^{2}\theta-aE)\right\}  \ ;\text{ 2nd-order}%
\end{Bmatrix}
, \label{eqn3.34}%
\end{equation}
where the functions $\widetilde{R}(r)$, $R(r)$, $\widetilde{\Theta}(\theta)$,
and $\Theta(\theta)$ are defined in Eqs. (\ref{eqn3.28})-(\ref{eqn3.29}),
while the auxiliary quantities and functions are described in Eqs.
(\ref{eqn3.15}) and (\ref{eqn3.17}).\footnote{Comparing the fourth-order
equations with the respective second-order equations of the standard theory,
it can be noted that the CG equations are obtained\ from the GR equations, by
performing the following substitutions: $M\rightarrow\widetilde{M}$,
$\Delta\rightarrow\widetilde{\Delta}_{r}$, $\Sigma\rightarrow\widetilde{\Sigma
}$, $\pm2Mr\rightarrow\mp\left[  \widetilde{\Delta}_{r}-(r^{2}-a^{2}%
)\widetilde{\Delta}_{\theta}\right]  $ and, in some appropriate places, by
replacing a factor of $1$ with $\widetilde{\Delta}_{\theta}$.}

It should be noted that our fourth-order equations of motion above are very
similar to the analogous equations of motion in Kerr-de Sitter spacetimes
(\cite{Kraniotis:2004cz}, \cite{Hackmann:2010zz}, \cite{Poudel:2013ura}), once
the CG parameter $k$ is connected to the cosmological constant $\Lambda$ by
setting $k=\Lambda/3$. In fact, our Eqs. (\ref{eqn3.31})-(\ref{eqn3.34}) have
the same structure, for example, of Eqs. (14)-(17) in Ref.
\cite{Hackmann:2010zz}, since the procedure used for the separation of
variables is the same. However, these two groups of equations differ in the
definition of the auxiliary functions, $\widetilde{\Delta}_{r}$,
$\widetilde{\Delta}_{\theta}$ and $\Delta_{r}$, $\Delta_{\theta}$,
respectively (and the additional function $\chi=1+\frac{a^{2}\Lambda}{3}$ is
used only in the Kerr-de Sitter case), so they are not completely equivalent.

As in the second-order case, following Eqs. (\ref{eqn3.31})-(\ref{eqn3.34}),
it is evident that the problem of solving the equations of geodesic motion has
been reduced to one of quadratures. In the next section, we will apply
directly our results to the particular case of the so-called \textit{Flyby
Anomaly} (FA) in order to check if our CG\ fourth-order solutions can explain
this gravitational puzzle.

As a final point of this section, we note that the new, fourth-order Carter's
constant $\widetilde{\mathcal{K}}$ (as well as the second-order $\mathcal{K}$)
can be obtained explicitly, in terms of $r$ or $\theta$, by combining together
Eqs. (\ref{eqn3.28}) and (\ref{eqn3.29}) with Eqs. (\ref{eqn3.31}) and
(\ref{eqn3.32}):%
\begin{align}
\widetilde{\mathcal{K}}  &  =-\widetilde{\Delta}_{r}p_{r}^{2}+\frac{\left[
\left(  r^{2}+a^{2}\right)  E-aL_{z}\right]  ^{2}}{\widetilde{\Delta}_{r}%
}-\delta_{1}r^{2}\label{eqn3.35}\\
&  =\widetilde{\Delta}_{\theta}p_{\theta}^{2}+\frac{(aE\sin\theta-L_{z}%
\csc\theta)^{2}}{\widetilde{\Delta}_{\theta}}+\delta_{1}a^{2}\cos^{2}%
\theta\ ;\text{ 4th-order}\nonumber\\
\mathcal{K}  &  =-\Delta p_{r}^{2}+\frac{\left[  \left(  r^{2}+a^{2}\right)
E-aL_{z}\right]  ^{2}}{\Delta}-\delta_{1}r^{2}\nonumber\\
&  =p_{\theta}^{2}+(aE\sin\theta-L_{z}\csc\theta)^{2}+\delta_{1}a^{2}\cos
^{2}\theta\ ;\text{ 2nd-order.}\nonumber
\end{align}
The alternative form of Carter's constant ($\widetilde{\mathcal{Q}}$ or
$\mathcal{Q}$) can be obtained using Eq. (\ref{eqn3.26}), or the equivalent
second-order relation.

\section{\label{sect:flyby}Geodesic motion and the Flyby Anomaly}

The Flyby Anomaly is a small unexpected increase in the geocentric range-rate
observed during Earth-flybys of some spacecraft (Galileo
\cite{Antreasian:1998a}, NEAR \cite{Antreasian:1998a}, Rosetta
\cite{Morley:2006}), as evidenced by both Doppler and ranging data. It is
usually reported as an anomalous change $\Delta V_{\infty}\sim1-10\
\operatorname{mm}%
/%
\operatorname{s}%
$ in the osculating hyperbolic excess velocity $V_{\infty}$ of the spacecraft
(see \cite{Anderson:2006tq}, \cite{Lammerzahl:2006ex}, \cite{Anderson:2009si},
\cite{Nieto:2009ve}, \cite{Turyshev:2009mt}, \cite{Anderson:2008zza} for
details and reviews), which is defined as:%
\begin{equation}
V_{\infty}=\sqrt{v^{2}(r,\theta)-\frac{2\mu}{r}}, \label{eqn4.1}%
\end{equation}
where $v(r,\theta)$ is the spacecraft speed along its trajectory, which will
be computed later in this section using the previous analysis of geodesic
motion. In the last equation $\mu=3.986004\times10^{20}\
\operatorname{cm}%
^{3}/%
\operatorname{s}%
^{2}$ is the universal gravitational constant times the Earth mass and $r$ is
the radial distance in a geocentric reference frame.

In Ref. \cite{Anderson:2008zza} an empirical formula was introduced which
could approximate well the observed anomalies in at least three cases (Galileo
first flyby, GLL-I; NEAR; and Rosetta) in terms of the incoming/outgoing
declinations ($\delta_{i}$ and $\delta_{o}$) of the asymptotic spacecraft
velocities. Explicitly:%
\begin{equation}
\frac{\Delta V_{\infty}}{V_{\infty}}=K\left(  \cos\delta_{i}-\cos\delta
_{o}\right)  , \label{eqn4.2}%
\end{equation}
with $K\equiv\frac{2\omega_{E}R_{E}}{c}=3.099\times10^{-6}$, where $\omega
_{E}$ and $R_{E}$ indicate respectively Earth's angular rotational velocity
and mean radius.

For three other Earth flybys (Galileo second flyby, GLL-II; Cassini; and
MESSENGER) it was either not possible to detect a clear anomaly, or any
anomaly at all. The recent Earth flyby of NASA's spacecraft Juno (October 9,
2013) is offering a new opportunity to observe the anomaly. While Juno's data
are currently being analyzed by NASA/JPL, preliminary studies
\cite{Iorio:2013lya} already rule out standard physics explanations of the
anomaly. Similarly, no convincing explanations exist for past flybys, coming
from both gravitational and non-gravitational physics (see
\cite{Lammerzahl:2006ex}, \cite{Turyshev:2009mt}, \cite{Iorio:2013lya} and
references therein).

Although it is still an open question whether the Kerr metric can be used for
rotating, axially-symmetric astrophysical objects other than black
holes,\footnote{The problem of finding a rotating perfect-fluid interior
solution, which can be matched to a Kerr exterior solution, has not been
solved yet.} we have nevertheless analyzed the geodesic motion in fourth-order
Kerr geometry associated with Earth flybys to check if the reported anomalies
can be explained by Conformal Gravity. In fact, the Kerr metric is a very good
approximation to rotating spacetimes, as long as they rotate slowly---which is
the case of Solar System applications (\cite{Shibata:1998xw},
\cite{Berti:2004ny}, \cite{Benhar:2005gi}). For other recent extensive studies
on geodesic motion and other solutions in conformal Weyl gravity see Refs.
\cite{Sultana:2012qp}, \cite{Said:2014lua}, \cite{Said:2012xt},
\cite{Said:2012pm}.

We also need to point out that, in view of the conformal equivalence between
the fourth-order Kerr metric and the second-order Kerr-de Sitter metric
(discussed in Sect. V of Ref. \cite{Mannheim:1990ya}) and of the similarities
between the equations of motion in these two geometries (as discussed in the
paragraph following Eq. (\ref{eqn3.34}) above), the geodesic motions in these
two cases are comparable. Therefore, our treatment of geodesic motion in
fourth-order Kerr geometry parallels in many respects the similar analyses of
geodesic motion in Kerr-de Sitter spacetimes (\cite{Kraniotis:2004cz},
\cite{Hackmann:2010zz}, \cite{Poudel:2013ura}).

First, we studied the general laws of geodesic motion in the fourth-order Kerr
geometry, following the similar analysis for the second-order case in Ref.
\cite{1992mtbh.book.....C} (Chapter 7, \S 63-64). In particular, we considered
the most general time-like geodesics by setting $\delta_{1}=1$ in previous
equations of Sect. \ref{sect:separability} (similarly, $\delta_{1}=0$ would be
used for null geodesics). We then considered the projection of the geodesics
on the $(r,\theta)$-plane:%
\begin{equation}
\int\nolimits_{r_{i}}^{r}\frac{dr}{\sqrt{\widetilde{R}(r)}}\mathcal{=\pm}%
\int\nolimits_{\theta_{i}}^{\theta}\frac{d\theta}{\sqrt{\widetilde{\Theta
}(\theta)}}, \label{eqn4.3}%
\end{equation}
which follows from Eqs. (\ref{eqn3.31})-(\ref{eqn3.32}). Here $r_{i}$ and
$\theta_{i}$ are some initial values for the two coordinates and a similar
equation holds also for the second-order case, with $R(r)$ and $\Theta
(\theta)$ instead of $\widetilde{R}(r)$\ and $\widetilde{\Theta}(\theta)$. The
double sign on the right-hand side of the equation allows for possible
increase/decrease of the radial distance with increasing (or
decreasing)\ values of the polar angle $\theta$.

It is well known that these orbits, even in the GR second-order case, are in
general non-planar, unless we restrict ourselves to purely equatorial orbits.
However, in the case being analyzed here (spacecraft in the Earth's
gravitational field), the classical unbound hyperbolic Newtonian orbits are a
very good approximation. Therefore, for each of the spacecraft motions being
analyzed, we used the reported Newtonian orbital parameters to calculate the
values of the constants of motion ($E$, $L_{z}$, $\mathcal{K}$, and
$\widetilde{\mathcal{K}}$), but then we computed the orbital motion using Eq.
(\ref{eqn4.3})---and the equivalent second-order orbit equation---in order to
seek possible discrepancies between the two cases.

It is customary to minimize the parameters by setting:%
\begin{equation}
\xi=L_{z}/E\ ;\ \eta=\mathcal{Q}/E^{2}\ ;\ \widetilde{\eta}%
=\widetilde{\mathcal{Q}}/E^{2} \label{eqn4.4}%
\end{equation}
and rewrite Eq. (\ref{eqn4.3}), and the equivalent second-order equation, in
terms of modified functions (in italics): $\mathit{R}(r)=R(r)/E^{2}$,
$\widetilde{\mathit{R}}(r)=\widetilde{R}(r)/E^{2}$, $\mathit{\Theta}%
(\theta)=\Theta(\theta)/E^{2}$, and $\widetilde{\mathit{\Theta}}%
(\theta)=\widetilde{\Theta}(\theta)/E^{2}$. The modified radial functions can
then be written explicitly as:%
\begin{align}
\mathit{R}(r)  &  =r^{4}+(a^{2}-\xi^{2}-\eta)r^{2}+2M[\eta+(\xi-a)^{2}%
]r-a^{2}\eta-r^{2}\Delta/E^{2}\label{eqn4.5}\\
\widetilde{\mathit{R}}(r)  &  =[1+k(\widetilde{\eta}+(\xi-a)^{2})]r^{4}%
+(a^{2}-\xi^{2}-\widetilde{\eta})r^{2}+2\widetilde{M}[\widetilde{\eta}%
+(\xi-a)^{2}]r-a^{2}\widetilde{\eta}-r^{2}\widetilde{\Delta}_{r}%
/E^{2},\nonumber
\end{align}
while the angular-$\theta$ functions and related integrals are more easily
computed in terms of $\mu=\cos\theta$ and using the additional parameter
$\alpha^{2}=a^{2}\left(  1-1/E^{2}\right)  >0$ for unbound orbits ($E^{2}>1$).

With some algebra the modified angular functions, expressed in terms of $\mu$,
become:%
\begin{align}
\mathit{\Theta}_{\mu}  &  \equiv\Theta_{\mu}/E^{2}=\eta-[\xi^{2}+\eta
-\alpha^{2}]\mu^{2}-\alpha^{2}\mu^{4}=\alpha^{2}[(\mu_{-}^{2}-\mu^{2})(\mu
_{+}^{2}-\mu^{2})]\label{eqn4.6}\\
\widetilde{\mathit{\Theta}}_{\mu}  &  \equiv\widetilde{\Theta}_{\mu}%
/E^{2}=\widetilde{\eta}-[\xi^{2}+\widetilde{\eta}-\alpha^{2}]\mu^{2}%
-[\alpha^{2}+ka^{2}(\widetilde{\eta}+(\xi-a)^{2})]\mu^{4}-ka^{2}[\alpha
^{2}-a^{2}]\mu^{6},\nonumber
\end{align}
where the first function (second-order GR case) can be expressed in terms of
the two constants%
\begin{equation}
\mu_{\pm}^{2}=\frac{1}{2\alpha^{2}}\left\{  [(\xi^{2}+\eta-\alpha^{2}%
)^{2}+4\alpha^{2}\eta]^{1/2}\mp(\xi^{2}+\eta-\alpha^{2})\right\}
\label{eqn4.7}%
\end{equation}
obtained by solving the quadratic equation associated with it: $\eta-[\xi
^{2}+\eta-\alpha^{2}]x-\alpha^{2}x^{2}=0$. Similarly, by solving the cubic
equation associated with the second function in Eq. (\ref{eqn4.6}),
$\widetilde{\eta}-[\xi^{2}+\widetilde{\eta}-\alpha^{2}]x-[\alpha^{2}%
+ka^{2}(\widetilde{\eta}+(\xi-a)^{2})]x^{2}-ka^{2}[\alpha^{2}-a^{2}]x^{3}=0$,
we could also factorize this function (fourth-order CG case) in terms of three
constants, $\mu_{1}^{2}$, $\mu_{2}^{2}$, and $\mu_{3}^{2}$, whose expressions
are rather cumbersome and will be omitted here.

The orbit equation (\ref{eqn4.3}) can be written in terms of the new functions
in Eqs. (\ref{eqn4.5})-(\ref{eqn4.6}), for both the GR and CG cases:%
\begin{align}
\int\nolimits_{r_{i}}^{r}\frac{dr}{\sqrt{\mathit{R}(r)}}  &  \mathcal{=}%
\mathcal{\mp}\int\nolimits_{\mu_{i}}^{\mu}\frac{d\mu}{\sqrt{\mathit{\Theta
}_{\mu}}}\label{eqn4.8}\\
\int\nolimits_{\widetilde{r}_{i}}^{\widetilde{r}}\frac{d\widetilde{r}}%
{\sqrt{\widetilde{\mathit{R}}(\widetilde{r})}}  &  \mathcal{=}\mathcal{\mp
}\int\nolimits_{\mu_{i}}^{\mu}\frac{d\mu}{\sqrt{\widetilde{\mathit{\Theta}%
}_{\mu}}},\nonumber
\end{align}
where, from now on, we will also distinguish between the radial coordinates
$r$, computed with second-order equations, and $\widetilde{r}$, computed with
fourth-order equations, since in general $r\neq\widetilde{r}$ for a given
value of the angle $\theta$ (or $\mu$).

In the angular integrals on the right-hand sides of Eq. (\ref{eqn4.8}), the
range of $\mu^{2}$ is between zero and the smaller of the two roots $\mu
_{-}^{2}$ and $\mu_{+}^{2}$ for the first equation, and between zero and the
smallest of the three roots $\mu_{1}^{2}$, $\mu_{2}^{2}$, and $\mu_{3}^{2}$
for the second equation.\footnote{This is actually true for the case of
unbound orbits with $\eta>0$ (or $\widetilde{\eta}>0$), which is valid for the
spacecraft motion related to the FA. See Ref. \cite{1992mtbh.book.....C}
(Chapter 7, \S 63-64) for a full discussion of all other possible cases.}
Calling $\mu_{\min}=\min(\left\vert \mu_{-}\right\vert ,\left\vert \mu
_{+}\right\vert )$ and $\widetilde{\mu}_{\min}=\min(\left\vert \mu
_{1}\right\vert ,\left\vert \mu_{2}\right\vert ,\left\vert \mu_{3}\right\vert
)$, in the two different cases, the range for $\mu$ is therefore $\mu\in
(-\mu_{\min},\mu_{\min})$ and $\mu\in(-\widetilde{\mu}_{\min},\widetilde{\mu
}_{\min})$, respectively, in the two lines of the last equation. The initial
radial value $r_{i}$ (or $\widetilde{r}_{i}$) corresponds to $\mu_{i}$, or
$\theta_{i}=\arccos\mu_{i}$, for both cases, where a convenient value for
$\mu_{i}$, or $\theta_{i}$, can be used. In this way, the resulting unbound
orbits will intersect the equatorial plane and be confined within the cones
$-\mu_{\min}<\mu<\mu_{\min}$ and $-\widetilde{\mu}_{\min}<\mu<\widetilde{\mu
}_{\min}$.

The integrals in Eq. (\ref{eqn4.8}) can be evaluated analytically in terms of
elliptic integrals of the first kind, or numerically with Mathematica
routines. Once the orbit equations in the $(r,\theta)$ and $(\widetilde{r}%
,\theta)$\ planes have been determined, at least numerically, the coordinate
velocities of a test particle undergoing geodesic motion can be computed by
combining the fundamental equations (\ref{eqn3.31})-(\ref{eqn3.34}) into the
following expressions:%
\begin{equation}%
\begin{Bmatrix}
\widetilde{v}_{r}\\
v_{r}%
\end{Bmatrix}
=\frac{\overset{\centerdot}{r}}{\overset{\centerdot}{t}}=%
\begin{Bmatrix}
\frac{\sqrt{\widetilde{R}(\widetilde{r})}}{\frac{\left(  \widetilde{r}%
^{2}+a^{2}\right)  \left[  \left(  \widetilde{r}^{2}+a^{2}\right)
E-aL_{z}\right]  }{\widetilde{\Delta}_{r}}+\frac{a\sin^{2}\theta(L_{z}\csc
^{2}\theta-aE)}{\widetilde{\Delta}_{\theta}}}\ ;\text{ 4th-order}\\
\frac{\sqrt{R(r)}}{\frac{\left(  r^{2}+a^{2}\right)  \left[  \left(
r^{2}+a^{2}\right)  E-aL_{z}\right]  }{\Delta}+a\sin^{2}\theta(L_{z}\csc
^{2}\theta-aE)}\ ;\text{ 2nd-order}%
\end{Bmatrix}
, \label{eqn4.9}%
\end{equation}%
\begin{equation}%
\begin{Bmatrix}
\widetilde{v}_{\theta}\\
v_{\theta}%
\end{Bmatrix}
=r\frac{\overset{\centerdot}{\theta}}{\overset{\centerdot}{t}}=%
\begin{Bmatrix}
\frac{\widetilde{r}\sqrt{\widetilde{\Theta}(\theta)}}{\frac{\left(
\widetilde{r}^{2}+a^{2}\right)  \left[  \left(  \widetilde{r}^{2}%
+a^{2}\right)  E-aL_{z}\right]  }{\widetilde{\Delta}_{r}}+\frac{a\sin
^{2}\theta(L_{z}\csc^{2}\theta-aE)}{\widetilde{\Delta}_{\theta}}}\ ;\text{
4th-order}\\
\frac{r\sqrt{\Theta(\theta)}}{\frac{\left(  r^{2}+a^{2}\right)  \left[
\left(  r^{2}+a^{2}\right)  E-aL_{z}\right]  }{\Delta}+a\sin^{2}\theta
(L_{z}\csc^{2}\theta-aE)}\ ;\text{ 2nd-order}%
\end{Bmatrix}
, \label{eqn4.10}%
\end{equation}%
\begin{equation}%
\begin{Bmatrix}
\widetilde{v}_{\phi}\\
v_{\phi}%
\end{Bmatrix}
=r\sin\theta\frac{\overset{\centerdot}{\phi\ }}{\overset{\centerdot}{t}}=%
\begin{Bmatrix}
\frac{\widetilde{r}\sin\theta\left\{  \frac{a\left[  \left(  \widetilde{r}%
^{2}+a^{2}\right)  E-aL_{z}\right]  }{\widetilde{\Delta}_{r}}+\frac{(L_{z}%
\csc^{2}\theta-aE)}{\widetilde{\Delta}_{\theta}}\right\}  }{\frac{\left(
\widetilde{r}^{2}+a^{2}\right)  \left[  \left(  \widetilde{r}^{2}%
+a^{2}\right)  E-aL_{z}\right]  }{\widetilde{\Delta}_{r}}+\frac{a\sin
^{2}\theta(L_{z}\csc^{2}\theta-aE)}{\widetilde{\Delta}_{\theta}}}\ ;\text{
4th-order}\\
\frac{r\sin\theta\left\{  \frac{a\left[  \left(  r^{2}+a^{2}\right)
E-aL_{z}\right]  }{\Delta}+(L_{z}\csc^{2}\theta-aE)\right\}  }{\frac{\left(
r^{2}+a^{2}\right)  \left[  \left(  r^{2}+a^{2}\right)  E-aL_{z}\right]
}{\Delta}+a\sin^{2}\theta(L_{z}\csc^{2}\theta-aE)}\ ;\text{ 2nd-order}%
\end{Bmatrix}
. \label{eqn4.11}%
\end{equation}

Therefore, the particle speed $v(r,\theta)=\sqrt{v_{r}^{2}+v_{\theta}%
^{2}+v_{\phi}^{2}}$, or $\widetilde{v}(\widetilde{r},\theta)=\sqrt
{\widetilde{v}_{r}^{2}+\widetilde{v}_{\theta}^{2}+\widetilde{v}_{\phi}^{2}}$,
for both cases, is obtained by combining the previous three equations with the
orbit equation $r=r(\theta)$, or $\widetilde{r}=\widetilde{r}(\theta)$,
computed from the integrals in Eq. (\ref{eqn4.8}). The osculating hyperbolic
excess velocity $V_{\infty}$ in Eq. (\ref{eqn4.1}) can be evaluated along the
particle orbit, for both the GR and CG cases, and possible differences between
the two situations can be postulated as a cause of the Flyby Anomaly. The
anomalous change $\Delta V_{\infty}$ in the hyperbolic excess velocity can be
due to the difference between the two values of $V_{\infty}$ computed with CG
and GR, i.e.:%
\begin{equation}
\Delta V_{\infty}=V_{\infty}^{(CG)}-V_{\infty}^{(GR)}. \label{eqn4.12}%
\end{equation}

Once the orbit equations, $r=r(\theta)$ and $\widetilde{r}=\widetilde{r}%
(\theta)$, have been computed, it is also possible---at least numerically---to
fully integrate the equations of motion (\ref{eqn3.31})-(\ref{eqn3.34}) and
determine all four coordinates $r$ ($\widetilde{r}$), $\theta$, $\phi$, and
$t$ as functions of the affine parameter $\tau$, or directly obtain the
spherical coordinates $r$ ($\widetilde{r}$), $\theta$, $\phi$ as functions of
time $t$, i.e., the classical equations of motion: $r=r(t)$, $\theta
=\theta(t)$, $\phi=\phi(t)$. However, this approach would be unnecessary for
the analysis of the FA in the context of the geodesics in the Kerr geometry
since the variations of $t$ and $\phi$ along the orbits do not reveal any
additional information about the velocities which is not already included in
the orbit equation. Therefore, we will not pursue the full integration of the
equations of motion in the following but only consider the orbit equations and
the related velocities.

Using the data reported in Table I of Ref. \cite{Anderson:2008zza} and those
retrieved directly from the HORIZONS Web-Interface by
JPL/NASA,\footnote{JPL/NASA website at: http://ssd.jpl.nasa.gov/horizons.cgi}
we have obtained the values of the constants of motion, $E$, $L_{z}$,
$\mathcal{K}$, and $\widetilde{\mathcal{K}}$, and computed the necessary
integration parameters ($r_{i}$, $\mu_{i}$, $\mu_{\min}$, $\widetilde{\mu
}_{\min}$) for the three flybys with a clear anomaly detection (NEAR, Rosetta,
GLL-I). Following the previous analysis, the orbit equation was numerically
integrated for different angular values of $\theta$ in order to obtain the
related radial coordinates $r$ - $\widetilde{r}$, and the respective
velocities along the orbits were computed with Eqs. (\ref{eqn4.9}%
)-(\ref{eqn4.11}). Our numerical routines also allowed us to change the two
main conformal parameters $\gamma\ $and $\kappa$ in order to study the effects
of Conformal Gravity on the orbits and velocities.

Using the values of $\gamma$ and $\kappa\ $in Eqs. (\ref{eqn2.11}%
)-(\ref{eqn2.12}), and standard values for Earth's mass, angular momentum,
etc., we obtained estimates of the anomalous change $\Delta V_{\infty}$ along
the orbits on the order of $\Delta V_{\infty}\sim10^{-8}-10^{-4}\
\operatorname{cm}%
/%
\operatorname{s}%
$ , thus negligible compared to the actual magnitude of the anomaly ($\Delta
V_{\infty}\sim0.1-1.0\
\operatorname{cm}%
/%
\operatorname{s}%
$). Larger values of $\Delta V_{\infty}$ can be obtained with our simulations
by increasing the conformal parameters.

In particular, since the effect of the $\gamma$ parameter, in the equations of
the previous sections, is simply to rescale the mass $M$ into $\widetilde{M}%
=M\left(  1-\frac{3}{2}\ M\gamma\right)  $, as in Eq. (\ref{eqn3.17}), its
variations do not affect the results considerably. On the contrary, the value
of the second parameter, $\kappa$, can affect the results significantly since
it is present in both $\widetilde{\Delta}_{r}$ and $\widetilde{\Delta}%
_{\theta}$\ terms.%

\begin{figure}[ptb]%
\centering
\fbox{\ifcase\msipdfoutput
\includegraphics[
width=\textwidth
]
{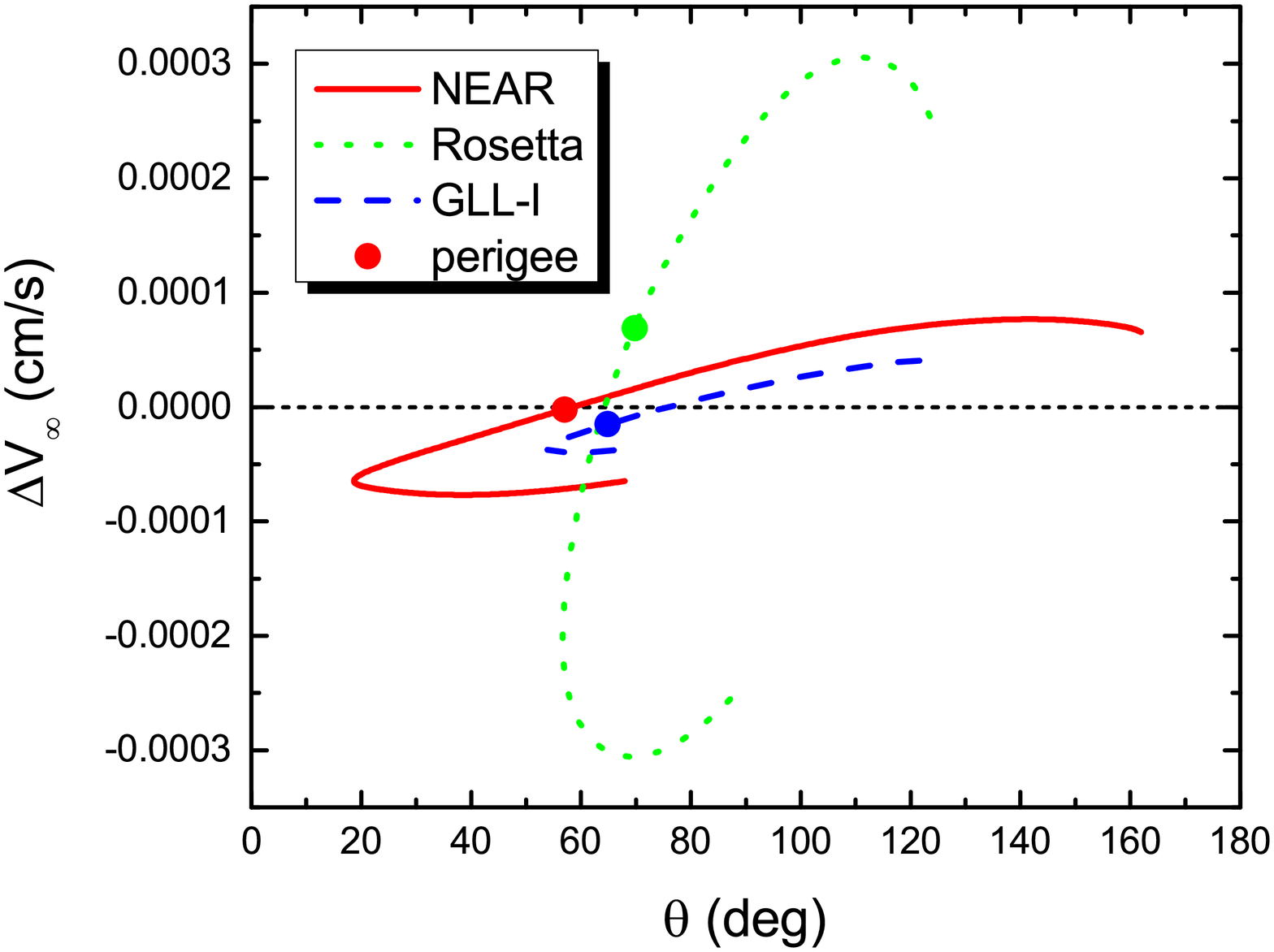}%
\else
\includegraphics[
natheight=3.135700in,
natwidth=4.068100in,
height=3.1357in,
width=4.0681in
]%
{C:/swp55/Docs/KINEMATICAL5/preprint/submit_arXiv_2/graphics/fig1__1.pdf}%
\fi
}\caption[CG results for the anomaly $\Delta V_{\infty}$ are shown here, for
the three cases examined, as a function of the polar angle $\theta$.]{CG
results for the anomaly $\Delta V_{\infty}$ are shown here, for the three
cases examined, as a function of the polar angle $\theta$. The computed
anomaly values are very small, compared to the observed values, due to the
choice of the second conformal parameter ($\kappa=6.42\times10^{-48}%
\operatorname{cm}^{-2}$). The simulation initial conditions were set to give
approximately zero anomaly at perigee angles (marked by solid circles in each
case) so that the anomaly values could be positive as well as negative in the
different parts of the orbits.}%
\label{fig1}%
\end{figure}

Therefore, we studied how the results for $\Delta V_{\infty}$ are affected by
possible variations of the parameter $\kappa$ for the three considered flybys.
Fig. 1 shows results of our numerical computations for values of the conformal
parameters as in Eq. (\ref{eqn2.12}), i.e., for $\gamma=1.94\times10^{-28}%
\operatorname{cm}%
^{-1}$ and$\ \kappa=6.42\times10^{-48}%
\operatorname{cm}%
^{-2}$. For each case, $\Delta V_{\infty}$ is computed as a function of the
polar angle $\theta$ (in degrees) along the spacecraft trajectory. For
example, in the case of the NEAR spacecraft (red solid curve in the figure),
the incoming declination $\delta_{i}=-20.76%
\operatorname{{{}^\circ}}%
$ corresponds to an incoming polar angle $\theta_{inc}=69.24%
\operatorname{{{}^\circ}}%
$, which decreases to a minimum angle of about $18%
\operatorname{{{}^\circ}}%
$, then increases to the perigee angle ($\theta_{per}=57%
\operatorname{{{}^\circ}}%
$, marked by a red solid circle in the figure), and finally reaches the
maximum outgoing angle $\theta_{out}=161.96%
\operatorname{{{}^\circ}}%
$, corresponding to the outgoing declination $\delta_{o}=-71.96%
\operatorname{{{}^\circ}}%
$. Similar plots are obtained for Rosetta (green dotted curve) and Galileo
GLL-I (blue dashed curve), where all the respective angles (incoming,
outgoing, minimum, perigee) were obtained from the data in Table I of Ref.
\cite{Anderson:2008zza}.

As seen in the figure, all the computed values for the anomaly are on the
order of $\Delta V_{\infty}\sim10^{-4}\
\operatorname{cm}%
/%
\operatorname{s}%
$, thus negligible compared to the experimental values reported. In this case,
the initial conditions for the integrals in Eq. (\ref{eqn4.8}) were set to
give approximately zero anomaly at the perigee angle (marked by solid circles
in each case) so that the resulting $\Delta V_{\infty}$ was negative during
the first part of the orbit (from the incoming angle to the perigee) and
positive during the second part (from perigee to the outgoing angle). It
should also be noted that our simulations show the largest anomaly for the
Rosetta case, while in the experimental data this spacecraft has the lowest
anomaly value of the three.%

\begin{figure}[ptb]%
\centering
\fbox{\ifcase\msipdfoutput
\includegraphics[
width=\textwidth
]
{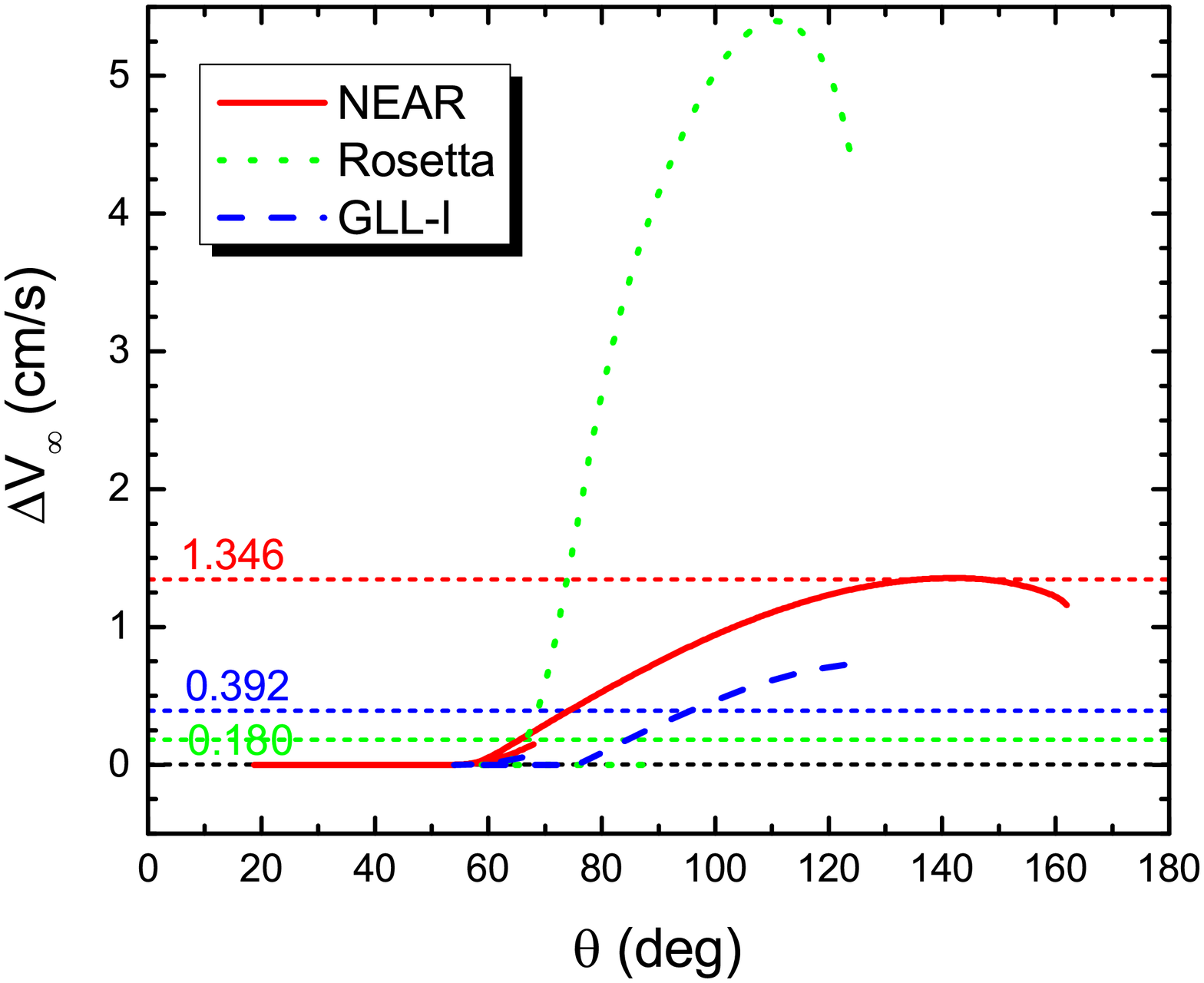}%
\else
\includegraphics[
natheight=3.136600in,
natwidth=4.068100in,
height=3.1366in,
width=4.0681in
]%
{C:/swp55/Docs/KINEMATICAL5/preprint/submit_arXiv_2/graphics/fig2__2.pdf}%
\fi
}\caption[CG results for the anomaly $\Delta V_{\infty}$ are shown here, for
the three cases examined, as a function of the polar angle $\theta$.]{CG
results for the anomaly $\Delta V_{\infty}$ are shown here, for the three
cases examined, as a function of the polar angle $\theta$. The computed
anomaly values are comparable to the observed values, due to an increased
value of the second conformal parameter ($\kappa\simeq5.00\times
10^{-40}\operatorname{cm}^{-2}$). The simulation initial conditions were set
to give approximately zero anomaly during the first part of each orbit, so
that the anomaly values are now restricted to positive range in the second
part of the orbits. Also shown are the experimental values of the anomaly
(short-dashed horizontal lines and related numerical values) for each case.}%
\label{fig2}%
\end{figure}

Fig. 2 shows instead results for an increased value of $\kappa\simeq
5.00\times10^{-40}%
\operatorname{cm}%
^{-2}$ (while keeping the same value as before for $\gamma$), thus obtaining
increased values for the anomaly $\Delta V_{\infty}$. The meanings of the
curves for the three spacecraft are similar to those of Fig. 1, but we have
adopted a different choice of the initial conditions for the integrals in Eq.
(\ref{eqn4.8}), which yields to almost zero anomaly during the first part of
the orbit and positive anomaly during the second part of the orbit. These
settings are similar to the original experimental fitting
(\cite{Antreasian:1998a}, \cite{Morley:2006}, \cite{Anderson:2008zza}) of the
Doppler and ranging data, where the pre-encounter trajectory was fitted to the
data (thus showing zero anomaly during the first part of the orbit), while
post-encounter data (second part of the orbit) were showing positive
residuals, therefore yielding positive values for the anomaly.

In Fig. 2, we also show the measured values of the anomaly, for the three
cases examined, represented by the horizontal short-dashed lines (from
\cite{Anderson:2008zza}:\ respectively,\ $\Delta V_{\infty}=1.346\
\operatorname{cm}%
/%
\operatorname{s}%
$ for NEAR, $\Delta V_{\infty}=0.180\
\operatorname{cm}%
/%
\operatorname{s}%
$ for Rosetta, $\Delta V_{\infty}=0.392\
\operatorname{cm}%
/%
\operatorname{s}%
$ for Galileo GLL-I). In this figure, the value of $\kappa\simeq
5.00\times10^{-40}%
\operatorname{cm}%
^{-2}$ was chosen to fit the maximum value of the computed anomaly for NEAR
(red solid curve) to the experimental value $\Delta V_{\infty}=1.346\
\operatorname{cm}%
/%
\operatorname{s}%
$ (red horizontal dashed line) since the NEAR\ anomaly is the most prominent.
As in Fig. 1, the computed Rosetta anomaly (green dotted curve) appears to be
much larger than the observed value, while the Galileo GLL-I anomaly (blue
dashed curve) is consistent with the value of the measured anomaly (blue
horizontal dashed line).

Therefore, our analysis shows that it is possible to obtain CG\ corrections to
the geodesic motion of spacecraft executing Earth flybys, yielding values for
the anomaly $\Delta V_{\infty}$ comparable to the observed ones, but only if
the value of the conformal parameter $\kappa$ is increased to about
$\kappa\sim10^{-40}%
\operatorname{cm}%
^{-2}$. Such a value is not supported by the current estimates of this
parameter in Eqs. (\ref{eqn2.11}) and (\ref{eqn2.12}), corresponding to a
range of $\kappa\sim10^{-53}-10^{-47}%
\operatorname{cm}%
^{-2}$. Our simulations also overestimate the value of the Rosetta anomaly,
compared to the NEAR and GLL-I cases.

If the $\kappa$ parameter is strictly constrained by cosmological data to the
above range of $\kappa\sim10^{-53}-10^{-47}%
\operatorname{cm}%
^{-2}$, then Conformal Gravity corrections to geodesic motion around Earth are
essentially negligible and are unlikely to be the origin of the Flyby Anomaly.
We leave further analysis of the FA, within the framework of CG, to future
work, once the data of the recent Juno flyby are made available.

\section{\label{sect:conclusions}Conclusions}

In this work we analyzed possible modifications to the Kerr geometry, geodesic
motion, and the problem of the separability of the Hamilton-Jacobi equation,
due to fourth-order Conformal Gravitational theory.

We obtained an explicit form of the equivalent Kerr metric in CG, expressed in
terms of the supplemental conformal parameters $\gamma$ and $\kappa$, thus
characterizing the geometry of a Kerr black hole in fourth-order gravity.

Using this explicit metric, we were able to show that the related
Hamilton-Jacobi equation is separable also in CG, and that an additional
conserved quantity---similar to the original Carter's constant---can be
introduced likewise in this case. As a consequence, geodesic motion in
fourth-order Kerr geometry can be studied along the lines of the similar
second-order GR case.

Assuming that the Kerr metric can be used as an exterior geometry for any
rotating axially-symmetric body, we have performed a limited analysis of the
geodesic motion of spacecraft executing flybys around Earth in order to assess
the interpretation of CG as the origin of the Flyby Anomaly.

Our preliminary analysis shows that CG\ is not likely to be the origin of the
FA, given the currently estimated values of the conformal parameters. However,
CG modifications of the geodesic motion might yield effects comparable to the
FA for increased values of the conformal parameters (in particular, of the
second parameter $\kappa$). Further studies will be needed to investigate this
possibility, in view of new FA data obtained from the recent Juno spacecraft flyby.

\begin{acknowledgments}
The author would like to thank Loyola Marymount University and the Seaver
College of Science and Engineering for continued support and for granting a
sabbatical leave of absence to the author, during which this work was
completed. The author is also indebted to Ms. Z. Burstein for helpful comments
and for proofreading the original manuscript.
\end{acknowledgments}

\bibliographystyle{apsrev}
\bibliography{CONFORMAL,FLYBY,MANNHEIM_RECENT,VARIESCHI}

\end{document}